# An Asynchronous Wireless Network for Capturing Event-Driven Data from Large Populations of Autonomous Sensors

Jihun Lee[1*], Ah-Hyoung Lee[1*], Vincent Leung[2], Farah Laiwalla[1], Miguel Angel Lopez-Gordo[3], Lawrence Larson[1], and Arto Nurmikko[1,4]

[1] School of Engineering, Brown University, Providence, RI, USA
[2] Electrical and Computer Engineering, Baylor University, Waco, TX, USA
[3] Department of Signal Theory, Telematics and Communications, University of Granada, Granada, Spain
[4] Carney Institute for Brain Science, Brown University, Providence, RI, USA

**Abstract**

We introduce a wireless RF network concept for capturing sparse event-driven data from large populations of spatially distributed autonomous microsensors, possibly numbered in the thousands. Each sensor is assumed to be a microchip capable of event detection in transforming time-varying inputs to spike trains. Inspired by the brain's information processing, we have developed a spectrally efficient, low-error rate asynchronous networking concept based on a code-division multiple access method. We characterize the network performance of several dozen submillimeter-size silicon microchips experimentally, complemented by larger scale *in silico* simulations. A comparison is made between different implementations of on-chip clocks. Testing the notion that spike-based wireless communication is naturally matched with downstream sensor population analysis by neuromorphic computing techniques, we then deploy a spiking neural network (SNN) machine learning model to decode data from eight thousand spiking neurons in the primate cortex for accurate prediction of hand movement in a cursor control task.

**Introduction**

A wireless network of spatially distributed radio-frequency identification (RFID) sensors can collect real-time information by streaming data to a single receiver from many nodes simultaneously [1-4]. Then, the received ensemble information can be used to make a prediction of the anticipated trajectory of a dynamically varying target environment in which the sensors are embedded. Such large-scale wireless sensor networks (WSN) composed of large populations of unobtrusive, battery-less, autonomous sensors are of contemporary interest for a wide range of potential applications including environmental sensing and health care [5, 6]. More broadly, those systems fall under the general heading of 'Internet of Things' (IoT) [7-9]. However, most commercial wireless RFID sensors available today lack a scalable network communication capability beyond a handful of devices. Therefore, one crucial issue for large-scale WSNs is the challenge of designing an efficient and robust wireless communication approach for rapid and low-error rate data retrieval; a problem for which solutions need examination outside the arsenal developed for mobile communications.

In our approach to a large-scale WSN, we take a cue from event-based, asynchronous, retina-mimicking dynamic vision sensor (DVS) cameras which convert time-varying illumination to sparse spike trains for high data transfer throughput and improved energy efficiency at each pixel [10-12]. We apply the concept to a network composed of remotely powered silicon chips which are assumed to be capable of converting recorded event data to time series of spikes. Then, we asked whether, in the case of event sparsity as with real biological neurons, one might be able to build a scalable wireless network of sensor "neurons" with improved efficient energy efficiency and bandwidth usage. However, well-established communication approaches such as versions of the random-access protocol, whether pure ALOHA, slotted ALOHA, or others [13-17], show severe limitations in the transmission of spike-type event data from larger populations of wireless RFID-type sensors. This follows from the requirement of fixed timeslots and/or bandwidth allocation in the conversion of sparse binary events into packetized data which undercuts the benefit of sparsity. That is, even when no meaningful events are detected, each node keeps transmitting and consuming valuable RF bandwidth. Alternative approaches based on impulse radio (IR) concepts have recently been used to address equivalent issues including the deployment of neuromorphic signal post-processing. But it appears to have considerable limitations, such as the inability to identify specific sensors, leading to limited scalability and data loss from the event of collisions [18, 19].

By contrast, a considerably more efficient and scalable WSN can be envisioned by transmitting binary spike events - after first digitally encoding the signal by a sequence of a pre-designed spreading code. This code serves as the address for a given asynchronous sensor node, a particular instance of a more general address event representation (AER) protocol [20], and allows an external receiver to recover and unpack spike event data from each sensor in the network. Multiplying data on-chip with a unique pseudo-random sequence (PN), such as the Gold or Kasami code, was theoretically analyzed in Refs 21 and 22 for an RFID tag. A related idea involving an asynchronous AER approach specific to neural sensing applications was recently proposed in optical communication domain using a pulse interval modulation scheme. The result of that work, based on computational simulations, suggested the possibility of transmission of up to 1,000-channel neuronal spike data via time-modulated LED light [23]. Other recent work includes that of He et al. who presented a single-sensor wireless neuromorphic sensing system as an integrated circuit and demonstrated compressive sensing and power-saving features in their chip [24]. In this case, low rate spiking data (from earthworms) encoding a peripheral neural waveform was transmitted using a 3-bit AER protocol for a single-point device.

In these and other prior studies, hardware demonstrations of multiple functional devices operating as a wireless network have been quite limited beyond a handful of nodes. And there has there been a lack of realistic assessment of the RF network configuration and the scalability of such a system. In this paper, we have focused both on the scalability of a WSN sensor system and a hardware proof-of-concept implementation of a communication chip designed as an Application-Specific Integrated Circuit (ASIC) as the path toward a realistic event-based microsensor network capable of acquiring data from thousands of asynchronous sensor nodes. In comparison with other wearable RFID sensors or related chip-scale RF sensors concepts [25-33], including our own recent work with RFID-type neural microsensors for brain implants [1, 34-35], the event-based RFID sensor network described below is shown to possess much larger scalability. Importantly, the 'Asynchronous Sparse Binary Identification Transmission' (ASBIT) network protocol, which we have developed, faithfully preserves timing information in the event detection and leads to an efficient means to transmit large amounts of event-driven data from a sensor population. Unlike many other communication schemes, the ASBIT method uses quasi-orthogonal codes, Gold code [21], for spike-event communication to minimize communication errors and significantly increase the capacity of a sensor network. Supported by experimental data with smaller populations of custom-designed silicon chips operating as a remotely powered, millimeter-scale, microwatt-power microchip network, we demonstrate in this paper the scalability of the ASBIT protocol *in silico* to thousands of sensor nodes at event error rates below $10^{-3}$. Additionally, we compare two different approaches to on-chip clocks in fabricated chips, necessary for on-board digital circuits, to optimize the network performance.

We then show how the population spike data received from such a wireless network lends itself naturally to decoding by neuromorphic computing techniques to predict the dynamics of a target environment. In particular, we asked whether the use of spiking neural network (SNN) based decoding methods might be effective in predicting state dynamics from spike-based ensemble recordings from the primate cortex. As a case example of direct relevance to current efforts to build high performance wireless brain-machine interfaces (BMIs) [36-39], we used the ASBIT network to transmit over 8,000 channels of spiking neural data recorded from the primate cortex for decoding. We customized an SNN algorithm to show a capability for accurate prediction of hand movement relative to the actual kinematics. While there are many other mathematical models used in computational neuroscience for decoding from populations of spiking neurons, ranging from the linear Kalman filter to convolutional neural networks, we conjecture that co-designing an ASBIT-type wireless networking approach with neuromorphic computing can be a low-latency, energy efficient means to build large scale WSN's for a range of applications from environmental sensing to healthcare.

**Overview of the proposed ASBIT communication method**

The proposed ASBIT networking approach is conceptually illustrated in Fig. 1a, which shows wireless transmission by ensembles of wireless microchip sensors that acquire their power from, and deliver backscattered signals to an external transceiver, here near 900 MHz. In this paper, we assume that each silicon chip includes circuitry which converts any time varying input signal to a spike train, not unlike the approach used in DVS cameras [10-12] (Fig. 1b). In an ASBIT protocol, sparse, asynchronous, binary data in the form of spikes is further encoded with a unique on-chip RF identifier, such as the Gold code, before transmission to a receiver through backscattering. Unlike conventional code-division multiple access (CDMA), the ASBIT scheme leverages statistical multiplexing of sparse data and information from event timing, analogous to a biological neuron firing [40, 41]. As a consequence, efficient use can be made of key network resources in terms of the spectrum, code, and timing (Fig. 1c).

We demonstrate below, combining benchtop experiments on populations of microfabricated sub-mm size CMOS chips with *in silico* simulations, how a single ASBIT link is scalable up to thousands of wireless sensors where the degree of event sparsity sets an error-rate limit for the system. The serial steps to unpack aggregate signals in the process of RF demodulation at the receiver are shown in the flow diagram of Fig. 1d. The raw incoming RF data (second panel from left) shows the example of a simulated aggregate signal from 1,000 sensors where the superposition of many backscattering signals masks information from any individual sensor. However, if data on each chip is digitally encoded with a unique identifier, such as the quasi-orthogonal Gold code described and implemented below, any set of binary events across the entire sensor ensemble can be recovered by an appropriate demodulation technique, here using matched filters (Supplementary Note 1, Supplementary Fig. 1). In the following, we first discuss the design of CMOS-based chips for testing the ASBIT protocol by explicitly considering two different methods for an on-chip clock to provide the timing for the on-board digital circuits. Namely, we compare the case of a free-running oscillator vs. the case where the timing signals are derived by frequency downconversion of the ~900 MHz baseband from an external transceiver unit. Further details of chip circuits and their characterization are given in Methods; relevant performance comparison between the two on-chip clock approaches is given in

Supplementary Note 3 and Supplementary Fig. 5. The microdevices in this paper employ the near-field (inductively coupled) electromagnetic regime for signal and power. However, the ASBIT-based communication approach is quite general and applicable to the far field or other communication modalities as well.

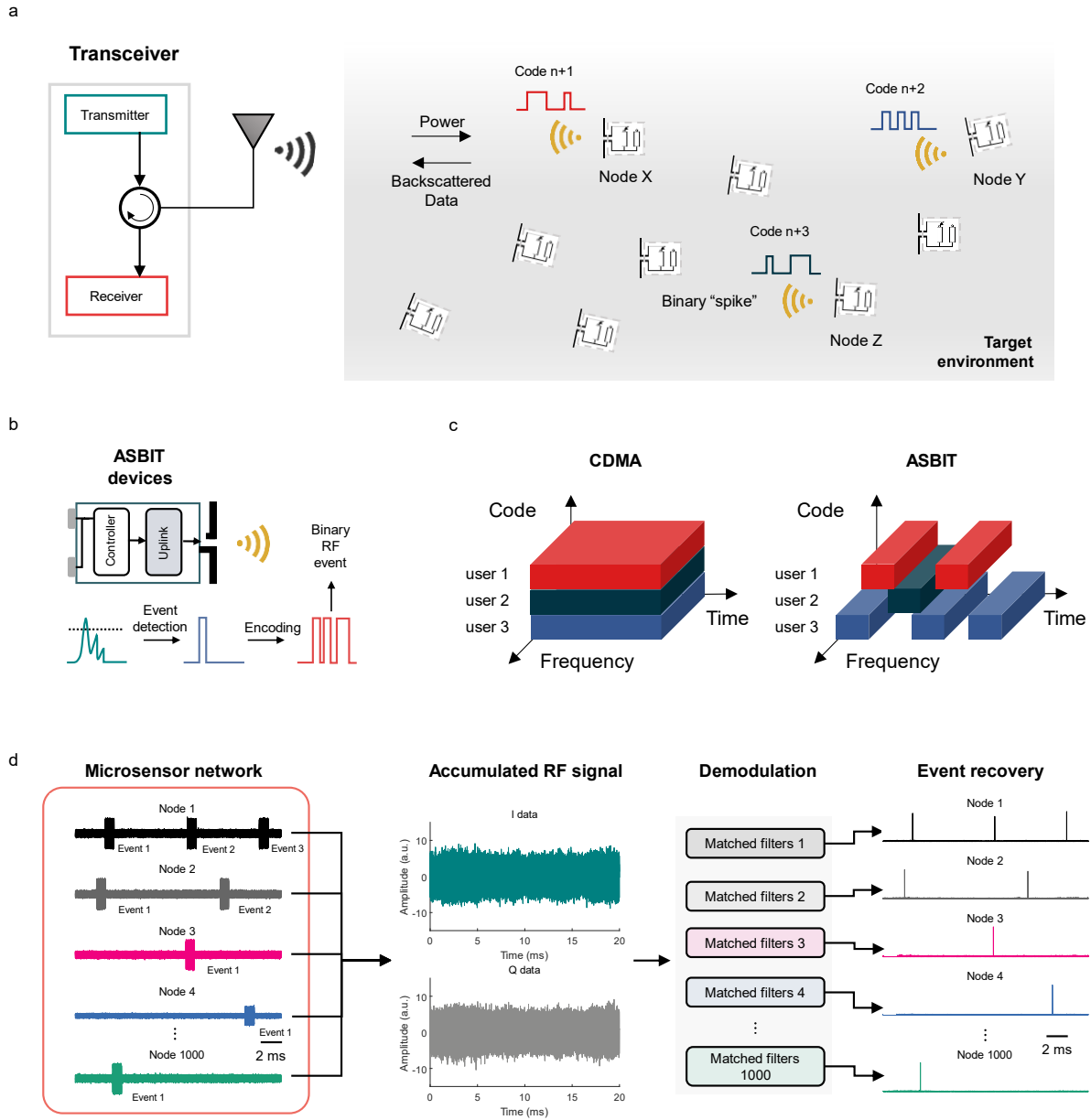

**Figure 1. a)** The concept of a 'detect and transmit' microsensor communication network (asynchronous, sparse, and binary). Each node transmits a detected sparse event via backscattering, encoded by a unique RF identifier sequence (e.g. a Gold code), which is captured by a transceiver antenna. **b)** Schematic illustration of a generic event detecting, neuronally inspired ASBIT sensor in which a binary signal is encoded with a pseudorandom code for RF transmission. **c)** Schematic highlighting the contrast in allocation of network resources (spectrum, code, and timing) by usual code division multiple access (CDMA) and by the ASBIT method, respectively. The latter requires uplink bandwidth only when reporting a detected event. **d)** A flowchart of illustrating the steps in unpacking the received ASBIT signals by demodulation (simulated data). Received IQ data (second panel from left) shows a raw superimposed signal from 1,000 nodes in this simulation, followed by the demodulation step where a battery of matched filters retrieves original events from each microsensor node.

**ASBIT microsensor network: the case of on-chip free-running oscillators**

The prototype wireless microchips with on-chip free-running clocks were designed to be sub-mm in size for potential use in body implants such as in our previous work on neural implants [1, 34, 35]. An overview of the circuit blocks in the ultra-low-power, system-on-chip silicon die is shown in Fig. 2a, including a low-voltage rectifier, a regulator, a oscillator, a generator for the Gold code address/identifier [21, 42], a digital finite state machine, plus a binary phase shift key (BPSK) modulator for backscattering (see also Supplementary Figure 1b,c for the chip layout). These communication chips were fabricated in TSMC's 65 nm mixed-signal/RF low-power CMOS process with the chip footprint miniaturized to a grain-of-sand size 300 µm × 300 µm (Fig. 2b). Constraints by small size and power as well as maintaining circuit simplicity make an onboard high-precision crystal oscillator or other advanced clock stabilization circuitry generally impractical. The simple free-running relaxation oscillator, we choose, had an operational frequency near 30 MHz.

The inset microphoto in Fig. 2b shows that a not insignificant fraction of the chip area is occupied by the multi-turn power harvesting coil around the chip perimeter plus the associated rectifier and impedance matching circuits. By way of comparison, the digital finite-state machine (FSM) which houses the communication code only requires an area of some 35 µm × 60 µm. For demonstration purposes, the communication circuit parameters were set to generate Gold-coded, backscattered transmission every 20 msec, corresponding to the case of an average event rate of 50 Hz. As a test of the receiver electronics, Supplementary Fig. 2 shows a clip of measured I/Q data from this wireless chip, the time window spanning the µsec timescale of a Gold code packet, compared with matched filter waveform generated by a custom synthesizer which we developed for this purpose (see Supplementary Note 1, Supplementary Fig. 3 and Methods for the details). As a practical matter, and to help make quantitative measurements of the wireless communication across a wide range of signal amplitudes for a population of chips, most of the benchtop tests were conducted using a three-coil near-field configuration in which a population of chips was encircled by an additional inductive relay coil in the same plane, separated by 10 mm from the transmitting (Tx) coil (photos in Fig. 2c) [1]. In RFID-type systems that use RF backscattering for communication, including our proposed ASBIT method, wireless transfer efficiency plays a central role in determining the signal-to-noise ratio (SNR). As the metric, we used the Received Signal Strength Indicator (RSSI), explained in detail in Supplementary Note 2 and Supplementary Fig. 4.

We fabricated a total of 78 wireless chips for the experiments, a practical number given limitations in a multipurpose wafer foundry run yet large enough for statistical performance analysis. We collected the backscattered signals from this population where the top left row of Fig. 2d shows the received aggregate signal. Because of the on-chip Gold code identifier (and its correlation properties), very good autocorrelation across the full chip population was verified for the received packets with no measurable communication errors, as expected for such a relatively small network (Fig. 2d, top right row). Then, to extrapolate the performance of the ASBIT protocol for much larger chip populations (nodes), we synthetically added "background signals", which are the previously collected data from actual chips, results being shown in the lower panels of Fig. 2d. To do this, we divided the chips into two groups, namely 40 "target" chips and 38 "background" chips and randomly added signals from the latter chips to artificially multiply the total population. For instance, when simulating 1,000 nodes, we selected each of the 38 chips 922 times, with the replacement, in a random sequence and added the packets from a given selected background chip to the original data (already containing backscattering data from the 78 chips). This method allowed us to analyze the communication fidelity within the target group while ensuring that no additional events were generated by the "target" chips themselves in assessing the contribution of a large number of nodes in the background. The examples in Fig. 2d of amplitude recovery from matched filter outputs demonstrate how the ASBIT method can differentiate a specific target packet even when multiple packets undergo interfering collisions. Even when 1000 nodes were included in the network whereby the original signal from a given chip is entirely obscured by the background, we were able to reliably detect the specific event in the output of the matched filter.

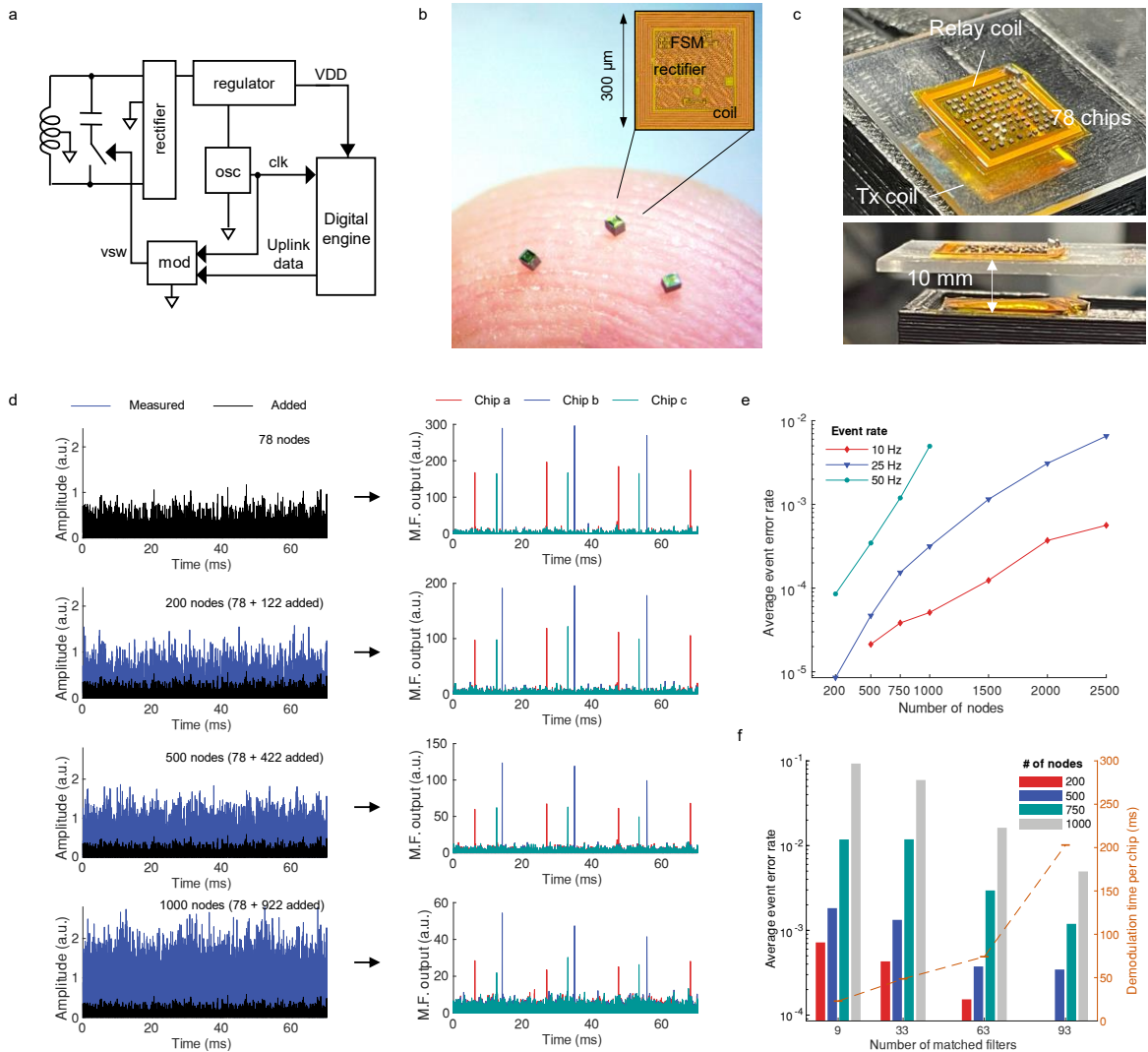

**Figure 2**. **Prototype of ASBIT system-on-chip wireless ASIC with on-chip oscillator, coil antenna, and digital logic circuit.** **a)** Principal circuit blocks of the ASIC with an on-chip antenna coil, rectifier, digital logic circuit, and oscillator as the system-on-chip clock. **b)** Microphotograph of wireless ASIC prototype chips mounted on a fingertip illustration of size, with inset showing the chip layout. **c)** Photograph of the benchtop 3-coil antenna configuration in wireless experiments (through 9 mm path in air, 1 mm glass); **d)** (Left) Combining measurements with simulations to study the scalability of the ASBIT network, with initial backscattered signals measured from 78 fabricated chips (here for an average SNR of 1.7 dB). A larger scale network was generated by simulating data backscattered from the equivalent of 200, 500, and 1,000 sensor nodes; (Right) Transient outputs of the matched filters (M.F.) for each case in d) (78, 200, 500, 1000 nodes). The outputs from the matched filters for data received is shown for three randomly chosen chips a,b, and c, obtained using three different sets of matched filters that were individually calibrated for each chip. **e)** Average EER (n=40, 6 second simulation epoch) as a function of the number of nodes added to the network. Each 'background' node detecting 10, 25, or 50 asynchronous events per second, respectively. Sparsity in event detection in the background enables a significantly larger number of nodes to operate in the network; **f)** Average EER as a function of the number of nodes and matched filters used for event recovery to account for the clock drift in each chip. The plot also shows how a longer demodulation time per node is required with increased number of matched filters (dashed line; 1 second data). Abbreviations: osc: oscillator, clk: clock, vsw: switching voltage, mod: modulator.

In the ASBIT concept, a bit "one" represents an event detected at a given sensor node, accompanied by the particular Gold code. To quantify the accuracy of data transmission, we solved for the Event Error Rate (EER) where a missing event or an instance of false detection was considered an error. Without loss of generality, we assumed a nominal duration of each event (bin size) of 1 msec, a choice appropriate for the neural application example in the last part of this paper. Fig. 2e shows the average EER of 40 target chips at an average SNR of 1.7 dB as a function of the number of other chips contributing to the background interference. We calculated that at the average event rate of 50 Hz, the ASBIT protocol can accommodate 750 nodes in achieving an EER of $1.19 \times 10^{-3}$, a value acceptable for many applications including for BMI use. We also tested the impact of different event rates for the background nodes to evaluate how sparsity affects the communication quality of the ASBIT protocol. Fig. 2e suggests that, without degradation of the EER, we can accommodate a significantly larger number of chips, up to 2,500, under sufficiently sparse events while demonstrating the protocol in utilizing sparsity without the need for any additional adjustments or further programming. The ability of each node to communicate with a single external receiver (or equivalently a single RFID reader) without the need for complex on-chip programming helps to keep the system architecture simple.

A penalty for using a miniaturized free-running oscillator in our microchip is it being subject to a clock drift of approximately ±1,000 ppm over time, the drift caused by fluctuations in chip voltage supply, ambient temperature, and other factors. The drift affects the waveform of the Gold code packet so that using a matched filter designed for a specific frequency can lead to inaccurate correlation values. To address this issue and to maintain the low EER values, we applied multiple sets of matched filters (Fig. 2f). The cost of using multiple matched filters is the increase in the demodulation burden as shown in Fig. 2f, which may be problematic applications requiring low latency. (For the description of the computational pipeline for sensor ensemble RF demodulation, please refer to the "Demodulation of ASBIT signals" in Methods).

**ASBIT microsensor network using an RF carrier frequency downconversion approach**

We next investigated the use of the RF downlink delivering power as a suitable frequency reference in circumstances where this might potentially be advantageous for our wireless sensor network concept. Using the baseband RF for timing has been demonstrated e.g. for RFID tags whereby the incoming RF frequency (in our case ~ 900 MHz) is divided by a fixed integer to generate a lower frequency clock [43-45]. One particular choice is a multiple-stage true single-phase clock (TSPC) frequency divider [46], shown schematically in Fig. 3a. The approach can offer the benefit of negligible clock variance and near-independence from energy harvesting efficiency, however at the expense of increased power consumption as depicted in Supplementary Note 3 and Supplementary Fig 5. Note that a frequency divider approach does not imply network synchrony since the phases of clocks of individual chips will differ due to phase lags arising e.g. from a random start-up in each chip's starting circuit.

We fabricated a set of chips in which the frequency divider circuit replaced the free-running oscillator as part of the same prototype ASBIT communication test chip. In benchtop measurements, we could obtain the timing sequences across the entire chip population as the demodulation step for the backscattered signals was now simpler in the absence of frequency shifts. In Fig. 3b, we illustrate how the application of matched filters within a predefined time window enables the direct exclusion of false detections, leading to an improved EER. One additional consequence is that the ASBIT protocol can be expected to accommodate even larger numbers of network nodes than in the case of an on-chip oscillator. We examined this network scalability by again using the method to synthetically add background signals (real chip) to the network as in the case of the free-running chip population above. Here, we used a slightly smaller population of 65 chips, the number limited due to post-processing loss during dicing and handling. The results in Fig. 3c then show how the ASBIT network can accommodate an ensemble of up to 2,500 microsensors while maintaining an EER of $1.03 \times 10^{-3}$ for an SNR of 1.7 dB (again assuming an average event detection rate of 50 Hz for each node). This represents more than a three-fold increase in comparison with the free-running on-chip oscillator case.

As a multidimensional summary of the EER analysis using the fabricated chip population, Fig. 3d shows how the network capacity is influenced by the average event rate across all nodes, here over a range of sensor nodes from 500 to 4,000. The heatmap was generated by multiplying the statistical event ('firing') rate by the number of nodes in evaluating the EER for the full network. As expected, the EER increases both with increasing event rate and the number of nodes. Overall, the aggregate event sum is what mainly determines the network communication performance and capacity. Here, approximately 100,000 events were collected per second, achieving an EER below $10^{-3}$. This result suggests that the ASBIT protocol can be adjusted to be quite flexible with a very simple scaling rule: A smaller number of sensor nodes allows for high event activity rates while a greater number constrains the network to sparser event rates.

Emulating real environmental effects, we further evaluated the effect of SNR on the ASBIT network performance by adding Gaussian white noise numerically to the experimentally acquired signals with the result shown in Fig. 3e. To account for the additional noise, we recalculated the average SNR by computing the ratio between the RSSI of each packet and the background noise level. As can be seen in Fig. 3e, the EER increases at low SNR values while the achievable error rate in a smaller network is considerably lower for the same SNR. For instance, for 1500 nodes an SNR of only -16.77 dB is needed to achieve an EER below $10^{-3}$. Thus, the ASBIT method can achieve acceptably low values of EER even when the backscattering amplitude is lower than our experimentally measured RSSI of -74.45 dBm as long as the number of nodes is limited to approximately one thousand range. The robustness to noise stems mainly from the effective bit protection provided by our Gold code strategy. For more information on coding gain, please refer to Supplementary Note 4.

Finally, we analyzed the ASBIT network by varying the length of the Gold code in simulations. Unsurprisingly, shorter codes offer lower coding gain and poorer auto- and cross-correlation properties. At the other end, very long codes can result in severe interference between backscattered signals which degrades the EER. Fig. 3f summarizes this relationship and shows that the optimal Gold code should be approximately 511 bits, the actual value we chose in our circuit co-design for the microfabricated chips. Longer code sequences also require more complex matched filtering thereby leading to an increase in demodulation time (see the dashed line curve in Fig 3f). Through simulations, we found that, for the case of a 511-bit Gold code, the demodulation time per a target node for a 1-second slice of data was only 12 msec, sufficient for many applications. As such, the drift-free frequency divider approach shortens the demodulation time as only three matched filters are needed to account for sampling the phases (see the comparison of the steps in the demodulation event recovery process for the on-chip oscillator and the frequency divider cases in Supplementary Fig. 6). In Fig. 3f, we synthesized waveforms for all packets due to the fixed length of the Gold code in the experimental chips, assuming an average SNR of 3.23 dB.

For practical applications, it is important to consider the design of wireless power transfer (WPT) for any large-scale battery-free sensor network. As a case example relevant to brain implants where RF energy in the near field (inductively coupled) regime must travel through a considerable thickness of lossy tissue (scalp, skull, and the dura), we show an approach using additional thin planar relay coils for enhanced WPT system in Supplementary Note 6 and Supplementary Fig. 9. On the other hand, given the scalability of the ASBIT method, the performance of a wireless network is not limited by data bandwidth in the first instance, especially for sparse events such as for typical neuronal firing rates. Rather, the number of sensors is likely limited by the WPT efficiency and the electromagnetic design of the energy delivery. When limited by regulatory considerations, e.g. specific absorption ratio (SAR), in increasing Tx power, it will be necessary to further reduce the sensor chip power consumption such as by transitioning to more advanced CMOS process nodes or utilize an additional power source to deliver extra energy [23].

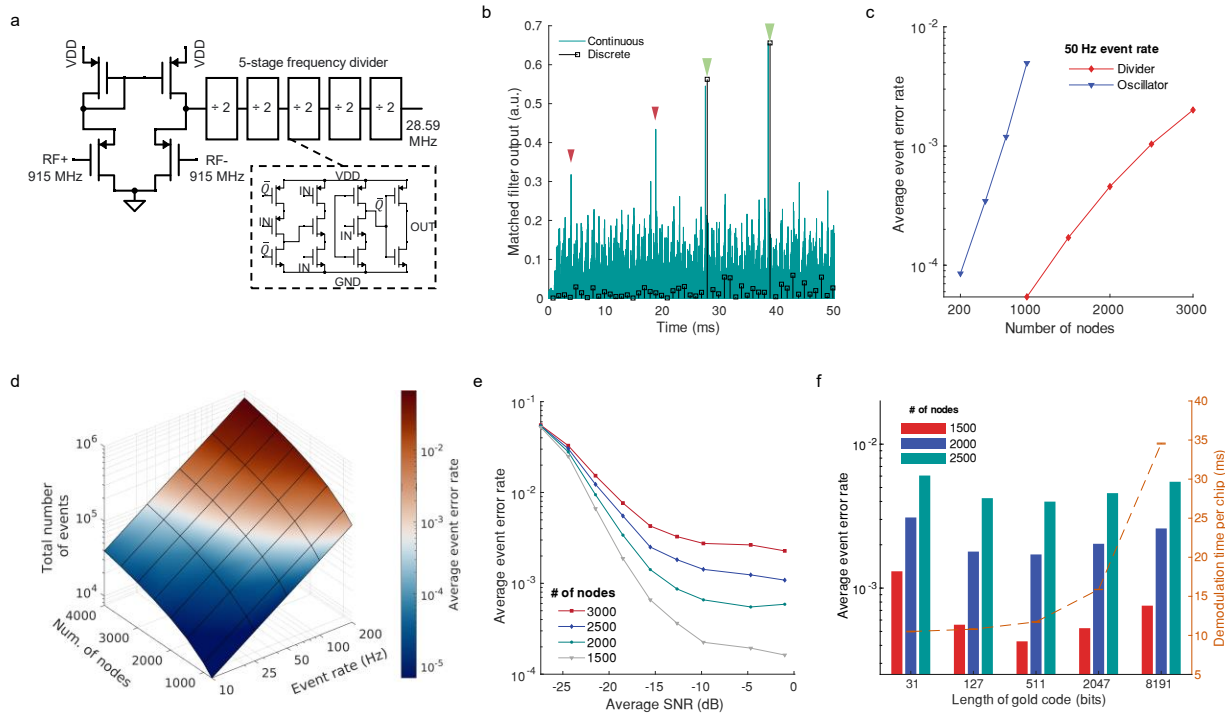

**Figure 3. An improvement of EER and network capacity for microchips with clock frequency divider. a)** Circuit diagram of a differential amplifier with multiple-stage true single-phase clock (TSPC) dividers to derive the clock frequency from a downlink RF energy source, in place of an on-chip oscillator. **b)** Comparison of matched filter (MF) outputs acquired from continuously received data in case of an on-chip oscillator vs. corresponding discrete RF data for chips incorporating a frequency divider for the clock (selected by estimating timeslots of a target wireless chip). The red triangles show instances that can cause false detection in this case of a continuous matched filter output while the green triangles indicate a correct event detection. **c)** Improved EER for the ASBIT chips with clock frequency divider, here the receiver is detecting 50 events per second over 6 seconds, as a function of the number of nodes in the network (average SNR of 1.7 dB). **d)** Heatmap representation of the overall network capacity, quantified as total number of transmitted events per second, and the averaged EER for selected 22 nodes (right hand color bar) as a function of the number of nodes and the event rate (SNR=1.7 dB, 30-second simulation time). **e)** Variation in average EER in relation to both network size and SNR, latter varied by adding Gaussian noise; 6-second simulation time. **f)** Dependence of average EER on the number of nodes and the length of the Gold code at SNR of 3.23 dB, for identifying an optimal code length for network performance. The plot also shows demodulation time per chip for 1 second piece of data, showing the increase when increasing the length of Gold of the code (average event rate is 50 Hz; transmission window is 6 seconds).

**A case example: Wireless transmission and population decoding by a spiking neural network (SNN) model of spike data from thousands of neurons in the primate cortex**

As an application example of the scalability of the ASBIT method, we considered the dual challenge for a mobile BMI to perform increasingly complex tasks: Wireless transmission of data recorded from thousands of points in the brain and the subsequent predictive neural decoding to operate an assistive device. The question can be posed more generally, namely as the co-design of means for data transmission from any sets of spatially distributed autonomous sensors embedded in a dynamical environment to the choice of the computing paradigm to predict its anticipated trajectory. We reasoned that, as an event-based method transmitting data in the form of spikes, the ASBIT technique might be well matched with large scale decoding tasks by the use of neuromorphic computing techniques. Neuromorphic computing, itself inspired by the brain, has recently developed to the point where low-power, portable hardware capable of executing sophisticated models of SNN at very low latency has become available [47]. At the same time, many mature methods exist in computational neuroscience for decoding neural data for BMI purposes, ranging from classical linear techniques (Kalman filter) to deep machine learning including our own recent work [48, 49].

Here, we assessed the scalability of the ASBIT network and determined the acceptable Spike Event Rate (SER) that can achieve the desired SNN neural decoding performance for hand velocity prediction in controlling a cursor. We used open source non-human primate data recorded from the primary motor cortex (M1) and its mapping on the somatosensory cortex (S1) [50, 51], both associated with hand movement intentions, to synthetically scale the spiking data to thousands of spiking neurons. In this context, Fig. 4a illustrates schematically how a population of hypothetical wireless microsensors detects neuronal spikes and transmits the spike ensemble events through the ASBIT protocol. The schematic follows the path of the signal flow from the sensors, implanted in or on the cortex, to the external SNN neural decoder. After demodulation of the Gold-coded RF data from a population of sensors, the decoder output is translated into hand-kinematic commands. The SNN decoder, described below and in Methods, accepted the binary spike data as the input and generated a reconstructed cursor velocity [52-55]. Since the neural signals themselves, the ASBIT protocol, and the SNN model are all spike-based, a seamless integration of these elements is both natural and efficient.

The firing of neuronal action potential spikes is inherently sparse [56, 57], shown in the example of Fig. 4b of raw data recorded in our laboratory from the auditory cortex of a non-human primate by intracortical microelectrode arrays [48]. Similarly, in Ref 50, the mean firing rate of 164 neurons in the primary motor cortex data during the center-out hand reaching task is shown as 9.2 Hz. Given this sparsity, we simulated a large-scale wireless BMI by including the wireless ASBIT protocol in the loop where the detected spike events are transmitted as Gold-encoded packets for each neuron. Since large-scale experimental neural data is currently unavailable, we used the limited open source data to synthesize equivalent spiking data from 8,200 separate nodes. Specifically, we transmitted 50 trial datasets from 164 electrodes simultaneously, as detailed in Supplementary Fig. 7. To generate the set of $50 \times 164 = 8200$ required Gold code packets, we developed a custom algorithm for synthesizing a matched filter which allowed for accurate modeling of the packet from chips as well (see Supplementary Note 1 and Supplementary Fig. 3).

We first analyzed the SER in ASBIT transmission, where - analogous to the EER - a missing target spike or a false detection counts as an error within a time bin of 1 msec. For this, we prepared an 8,200 M1 neural dataset for a 24-second and an 8,320-channel S1 data set for a 19-second data acquisition window. We calculated the average SER of the received spike trains across the 8000-plus sets of nodes as a function of SNR with the result in Fig. 4c. Assuming an average SNR= -0.77 dB, not unrepresentative of animal experimental situations, we computed an average SER of $3.23 \times 10^{-4}$ in M1 data and $6.12 \times 10^{-4}$ in S1 data for the entire network. Meanwhile, to emulate the near-far electromagnetics problem in an implant, where a given microchip closer to the Tx coil generates a stronger signal, we varied the backscattering amplitude by up to a 20 dB difference in SNR across sensor microchips. Fig. 4d and Supplementary Fig. 8 show the dependence of the average or individual node's SER, respectively, on the relative SNR, indicating that the backscattered amplitude of a specific node dominantly determines the SER of that node. This should be similar in far-field or other regimes where the external power source projects a spatially inhomogeneous electric or magnetic field.

For the neuromorphic deep-learning approach, we trained an SNN model to reconstruct the cursor velocities using the movement-intention evoked intracortical spike data. And we compared the outcomes for the hypothetical case of ideal, lossless wired electrodes (data also synthesized from the open source data) with the case where such data was transmitted through the wireless ASBIT chain. Fig. 4e depicts the relationship between the actual kinematic x-component of the hand velocity toward a target and the reconstructed velocity using cortical spike data. The top left plot presents the prediction using neural data without any spike loss showing a linear relationship between the kinematic and the reconstructed velocities. The other plots compare the kinematic velocity with the prediction obtained from spike data transmitted through the ASBIT protocol for different SNRs affecting the number of spike errors. In the low SNR regime, the data points deviate from the ideal line and the decoding performance degrades.

To more quantify the decoding performance, we computed the correlation coefficient between the original velocity and the reconstructed velocity on testing sets [50, 58], as described in the Methods section. We obtained the correlation coefficient using the wired neural spike signals, referred to as $r_{neural}$, and using the wirelessly transmitted neural signals, denoted as $r_{SER}$. Since the ASBIT link may introduce spike errors, we expect $r_{SER}$ to be lower than $r_{neural}$. In the absence of spike errors, average $r_{neural}$ were 0.9324 for M1 data and 0.9297 for S1 data, both of which were obtained using a 5-fold approach. All correlation values obtained in the 5-fold cross-validation are presented in Supplementary Table 2. Fig. 4f shows the ratio of $r_{neural}/r_{SER}$ as a function of the SER. The results suggest that, even with an SER of $10^{-4}$, the decoding performance of the ASBIT system ($r_{SER}$) is comparable to that achieved by wired microelectrode arrays ($r_{neural}$), as $r_{SER}/r_{neural}$ approaching unity. Even for an SER of $10^{-3}$, the wireless ASBIT link can still achieve 88.8% and 85.4% of the decoding performance compared to an equivalent wired system for M1 and S1 data, respectively. Note that we assumed the simultaneously transmitted 50 datasets for M1 as well as the 160 datasets for S1 to be independent of each other. Yet, since many neurons in the primate brain encode shared information and were not independent of each other, we anticipate that a large-scale ASBIT-based wireless neural interface, when implemented as a chronic implant, is likely to have better decoding performance, $r_{neural}$, in being more robust to spike errors.

We also conducted an additional neural decoding experiment to investigate the importance of preserving high-resolution spike timing information through the ASBIT protocol. We first introduced random jitter to the spike timing (up to $\pm$ 25 ms) in each SNN training epoch of recorded neural data transmitted via the wireless link. The random jitter was introduced to mimic the stochastic character of neurons and to enable the SNN model to be trained across many different temporal patterns of spike trains. We then compared the correlation values with and without jittering, as shown in Fig. 4g. We performed Fisher Z transformation on the correlation for statistical analysis and found that the average correlation was in fact higher with spike jittering for both M1 and S1 data. A paired t-test showed a statistically significant difference between the two cases in both datasets (p = 0.0033 in M1, p = 0.00007 in S1). These results suggest that preserving spike timing information at high resolution is important in building an SNN-based neural decoder that is robust to spike errors and inherent stochastic activity in neurons, thereby leading to improved decoding performance.

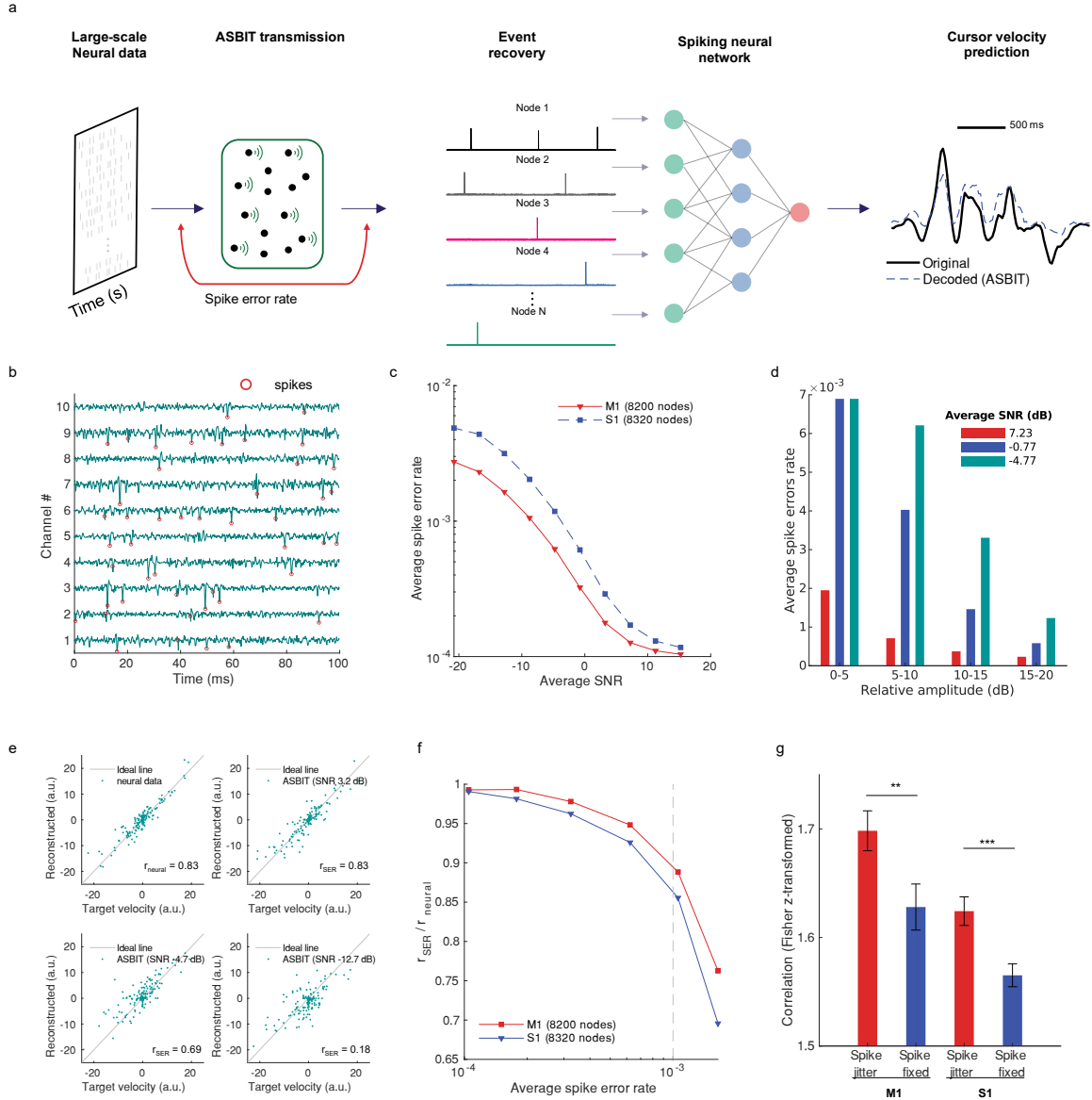

**Figure 4. Transmitting and decoding over 8000 channels of spiking neural data. a)** Schematic illustration of the wireless transmission of neuronal spike events using the ASBIT protocol. Received signal at an external wireless hub is demodulated to extract spike data from each node as input to a SNN-based decoder to predict cursor velocity. **b)** Example of raw broadband data from auditory cortex [48], sampled at 30 KSa s-1 from ten channels. The red circle shows neural spikes detected by thresholding. **c)** Spike error rate vs. SNR, latter controlled by adding Gaussian white noise to M1 data (8200 channels, 164 neurons × 60 datasets) and S1 data (8320 channels, 52 neurons × 160 datasets). Large-scale neural data was synthesized as described in Supplementary Fig. 7. Data obtained from [50, 51]. **d)** Average spike error rates computed based on the backscattering amplitude indicate that the average SNR and amplitude are the two dominant factors affecting SER. **e)** The relationship between target x velocity and reconstructed x velocity from the SNN neural decoder. The top-left graph shows the reconstructed velocity using neural data collected from a wired system, while the other graphs show the decoded velocity using spike data transmitted through the ASBIT protocol at different SNRs (200 data points). The gray line represents the ideal line where the target and reconstructed velocity exactly match. **f)** Dependence of neural decoder performance on average SER showing correlation between cursor velocity and velocity reconstructed by using spike data transmitted through the ASBIT ($r_{SER}$). The value for $r_{SER}$ is normalized to correlation obtained by using original spike data ($r_{neural}$). Each correlation value is averaged across five folds. **g)** Effect of spike timing jitter on the neural decoder's performance, highlighting the advantage of transmitting neural spikes to preserve precise spike timing information (here SNR=3.23 dB). For statistical analysis, correlation values are obtained in tenfold cross-validation [60] and Fisher Z-transformed to approximate the normal distribution. Error bars are presented as mean values ± s.e. (n=30, repeated ten-fold cross-validation, three trials). Abbreviation: M1: primary motor cortex, S1: primary sensory cortex.

## Conclusion

We have proposed a communication approach for large-scale wireless asynchronous microsensor networks to transmit binary events from thousands of local nodes with spectral efficiency and at low error rates. The received data serves as input for the reconstruction of the dynamical states of the environment in which the sensors are inserted. The ASBIT approach, with each node reporting binary events above the threshold, is inspired by principles governing information processing in the brain including intrinsic sparsity of single neuron activity. We have demonstrated hardware proof-of-concept in implementing the ASBIT protocol on submillimeter-sized silicon chips and showed low error rate transmission and robustness from collisional interference, deploying an event recovery method based on matched filter techniques. We have explored two different on-chip timing strategies, a free-running oscillator, and a clock frequency divider. The former allows low-power autonomous operation of microsensors whereas the latter enables further system scalability in reducing transmission error rates and latency at the expense of higher power consumption. A description of how our method compares with other relevant communication techniques is given in Supplementary Note 5 and Supplementary Table 1. Lastly, we provided a simulation example of the scalability of an ASBIT network in the joint BMI context of spike-based wireless transmission and SNN model neural decoding. We used sets of experimental neural data recorded from the primate cortex, synthetically scaling up to 8,320 independent spiking channels, in decoding cortical state dynamics for the reconstruction of cursor movement intention. More generally, the work in this paper should be applicable to the design of other large-scale wireless sensor networks where, using a spike-based binary approach, the transmission at low error rates combined with neuromorphic data analysis can be useful in predicting the dynamics of a complex heterogeneous target environment of interest.

## Methods

### The ASIC chip design and functional validation

The prototype microchips which were designed to operate in the ASBIT transmission mode were designed by partly leveraging our previous work in using the TSMC 65 nm mixed-signal/RF low-power CMOS foundry process. A diode clamp provided over-voltage protection from the energy harvesting process for the circuits while a low-dropout regulator helped to stabilize the voltage supply for the on-chip oscillator. Gold code generator and digital processor were implemented on a finite state machine (FSM), running at 1/3 speed either of the free-running oscillator or frequency divider output, respectively, (nominally at 10 MHz). We tested the silicon chips received from the foundry under fully wireless conditions in the laboratory using a software defined radio (the SDR Model 'Raptor' by Rincon Research or FMCOMMS3 by Analog Devices) and a single-turn polyimide printed circuit board which included the transmitter Tx coil and relay coil (both 9 mm × 9 mm area). The SDR, together with Analog Devices AD9361 transceiver chips and Zynq SoCs, generated an RF baseband carrier at 915 MHz, which was further amplified by an RF power amplifier (ADL5605-EVALZ, Analog Devices). An RF surface acoustic wave duplexer (D5DA942M5K2S2, Taiyo Yuden) isolated the backscattered signals from the downlink carrier. The SDR performed amplification of backscattered signals, downconversion from 945 MHz to DC, and finally the analog-to-digital conversion at 30 MSa s$^{-1}$ (12-bit). (as in Supplementary Fig. 2). The digitized IQ data were then ported to a personal computer for offline ASBIT demodulation (computation in MATLAB/Simulink).

**Modeling of backscattering signals from RFID microsensors for matched filter design**

Equations (1)-(6) and Supplementary Fig. 3 illustrate a stepwise simulation process of RF signals from a microsensor system utilizing backscattering modulation where the RF transceiver hub emits a continuous carrier wave as

$$x_c(t) = Re[A_{Tx} e^{j2\pi f_{Tx} t}] \quad (1)$$

where $A_{Tx}$ and $f_{Tx}$ denote the amplitude and the frequency of the carrier (Tx), respectively. The incident backscattering waves can be expressed as

$$s_{bck}(t) = \sum_{n=1}^{N_s} S_n p_1(t - nT_s) \quad (2)$$

$$\text{where, } p_1(x) = \begin{cases} 1 & \text{if } -T_s/2 \le x \le T_s/2 \\ 0 & \text{otherwise} \end{cases}$$

$S_n$ is the coded symbol sequence that takes either the value +1 or -1, $N_s$ is the number of symbols in the sequence, and $T_s$ is the duration of a symbol. In the proposed ASBIT protocol, the symbols are encoded in BPSK modulation using a system clock ($f_{clk}$) which is three times faster than the symbol rate [1], thus $f_{clk} = 3/T_s$. Therefore, $s_{BPSK-bck}$ is defined as

$$s_{BPSK-bck}(t) = \sum_{n=1}^{N_s} \sum_{m=1}^{3N_s} S_n p_1(t - nT_s) \cdot S_{clk} p_2(t - mT_s/3) \quad (3)$$

$$\text{where, } p_2(x) = \begin{cases} 1 & \text{if } -T_s/6 \le x \le T_s/6 \\ 0 & \text{otherwise} \end{cases}$$

$S_{clk}$ is the clock sequence, which also takes either the value +1 or -1. In our case, a chip generates backscattered data by modulating its antenna impedance between two states and the modulated backscattering signal can be approximated as

$$x_{bck} = Re[A_{bck} \cdot s_{BPSK-bck}(t) \cdot e^{j(2\pi f_{Tx} t + \phi)}] \quad (4)$$

Here, $A_{bck}$ is the backscattered amplitude determined by modulation depth and backscattering cross-sections, and $\phi = \frac{2\pi}{\lambda} D$ is the phase delay [21]. $\lambda$ and D denote the wavelength and the distance between the transceiver hub and the ASBIT chip, respectively. A duplexer at the transceiver isolates $x_{bck}$ from the downlink carrier from the software radio (SDR) capturing the backscattering signal. After a low-noise amplifier (LNA), a downconversion with a conversion factor $\alpha'$ is performed at $f_{Tx} + f_{n.clk}$ (nominal clock frequency, 30 MHz) to yield

$$y(t) = \alpha' G A_{bck} s_{BPSK-bck}(t) \cdot \{\cos(2\pi f_{n.clk} t - \phi) - j \sin(2\pi f_{n.clk} t - \phi)\} + \omega(t) \quad (5)$$

where G is the LNA gain in the SDR and the noise signal ω(t) is modeled as a zero-mean complex additive white Gaussian noise (AWGN) process. The received signal is then passed through an ADC, whose output sequence $y(v)$ can be defined as

$$y(v) = A' s_{BPSK-bck}(v) \cdot \{\cos(2\pi f_{n.clk} v + \phi') - j \sin(2\pi f_{n.clk} v + \phi')\} + \omega(v) \quad (6)$$

$$\text{where, } v = 1, \dots, N_{sc} N_S$$

which is the received signal sampled at time instants $t = v T_s/N_{sc}$. Here, $N_{sc}$ is the number of ADC samples per coded symbol and $A' = \alpha' G A_{bck}$. Due to asynchronous properties in our network, the ADC sampling also can affect the phase of the received signal, thus $\phi' = \phi + \phi_{ADC}$ where $\phi_{ADC}$ is the sampling phase.

According to Eq. (6), the received signal is equivalent to the down-converted $s_{BPSK-bck}(v)$ with an amplitude of $A'$ and the phase shift of $\phi'$. We performed downconversion on $s_{BPSK-bck}(v)$ to model the received signal in our simulation while also including various amplitudes $A'$, clock frequencies $f_{clk}$, phase shifts $\phi'$, which collaboratively determine the waveform of received signals. Also, this down-converted $s_{BPSK-bck}(v)$ is directly employed as a matched filter during the demodulation step. When simulating the presence of multiple nodes (N microsensors in the network), the received signal $Y(t)$ from the entire network was modeled as a superposition $y_i(t - \tau_i)$ where $i = 1, \ldots, N$ and $\tau_i$ denotes the initial delay time of backscattering.

**Encoding using the Gold code**

In the ASBIT protocol, each sensor chip has its unique Gold code identifier, i.e. one from the many random sequences often called "pseudo-noise" (PN) sequences. Contrary to this designation, the Gold code is however predictable and reproducible. Among such PN sequences, Maximum Length (ML) codes are the largest codes that can be generated by a shift register and have a period of $2^m - 1$ where m is the length of the shift register. Gold codes are generated from XOR multiplication of a preferred pair of ML codes and, in given m, the set of Gold codes is made up of $2^m + 1$ codes in a total of length $2^m - 1$. Gold codes are suitable for spread spectrum systems and are often characterized by their auto- and cross-correlations. The selection of the length of the Gold code is important because the length sets bounds on network capacity and its auto- and cross-correlation properties. The microsensors in the ASBIT network generate an $S_n$ using the Gold code generator in the on-chip digital FSM which is encoded into BPSK hence yielding $s_{BPSK-bck}(t)$. This is the $s_{BPSK-bck}(t)$ which is backscattered whenever the sensor front end detects an event.

The PUF was generated by intrinsic 65 nm CMOS fabrication process variations and unique to each chip (see also Supplementary Fig 1). We implemented a circuit design that utilized a 13-bit PUF as the seed for our transmitted 511-bit Gold code sequence. By allowing each chip to synthesize up to 8191 bits of Gold code, we were able to increase scalability and reduce the chance of PUF collision. However, to reduce mutual interference and achieve better performance in larger networks, we used only 511 bits out of the 8191 bits Gold code sequences. This decision was based on our analysis of the optimal length of the Gold code as described in the text.

**Demodulation of ASBIT signals**

In order to demodulate the received ASBIT coded data stream, summarized in Supplementary Fig. 6, we used multiple sets of matched filters for the received signal $Y(t)$. A set of matched filters for the target node was designed by using the pre-discovered digital Gold code accounting for the phase shift $\phi'$, the clock frequency and the clock drift over time. Using those parameters, we synthesized multiple functions $y_i(t)$ following equation (5) and directly used these waveforms as matched filters. Note that we needed to separately build matched filters for I data and Q data and then multiplied these two at each time point to build a whole robust to phase cancellation. The first step of the demodulation was the clock recovery process where we synthesized matched filters assuming various values of $f_{clk}$, ranging from 27 – 33 MHz and applied them to short clips of incoming data $Y(t)$ to check the matched filter output dependence on $f_{clk}$. By checking the maximum value of a matched filter output, we could find whether the target chip generated backscattering pulses (or not) and could thus solve its clock frequency. Using the discovered clock frequency, we synthesized a smaller set of matched filters although clock drift still needed to be accounted for. For example, for clock drift of 1,005 ppm (± 30 kHz), we used a total of 93 matched filters (3 phase variants × 31 clock points). In case of the frequency divider scenario, on the other hand, one only needed up to 3 matched filters for correcting phase variants as the clock frequency was now well-defined (zero drift assumed). In order to accelerate the demodulation process, we can first find the exact time slot of a given node and solve $\tau_i$, then apply matched filters to signal function $Y(t)$ by a discrete timing approach (which can be done in a phase recovery process also by using a short piece of $Y(t)$). After synthesizing matched filters, we applied these to $Y(t)$ to get multiple matched filter outputs which were summed to get their combined continuous or discrete matched filter outputs. Finally, events reported by any sensor node could be detected by thresholding the combined matched filter output. We determined the threshold based on the root mean square of the matched filter output (which makes the process adaptive to the properties of the network).

**Decoding of primate neuronal population data using recurrent spiking neural networks**

We evaluated the performance of population neural decoding using spiking neural network (SNN) algorithm from the Lava-dl library [52, 59] and assessed how the transmission fidelity of neural spiking signals in the ASBIT protocol affected the neural decoding. We chose an SNN technique as opposed to another machine learning approach based on intuition outlined in the text but also as can be readily implemented on recently developed portable Loihi2 neuromorphic hardware for real-time processing [47] suitable in an BMI application.

Our SNN model comprises of three dense neural network layers and a sigma-delta neuron input and output layer. The model used 650 ms of spike data at 1 ms resolution to estimate the velocity of the cursor in the x-direction. To account for the inherent stochastic nature of neural activity, we randomly shifted the spike times (up to ± 25 ms) during each training epoch, a process known as spike jittering. After spike jittering, we binned the spike data into non-overlapping 25 ms intervals to improve the model's robustness to variations in neural activity. We trained all models for 200 epochs using Adam as the optimization algorithm and monitored the event rate loss metric. We split the dataset, obtained from [50, 51], into training and testing sets made up of 80 %, 20 % the data in 5-fold cross validation. In repeated 10-fold steps of cross validation [60], we split the data into 90 % training sets and 10% testing sets. We trained the SNN model with training data while the testing set was only used to assess that final performance of the model in predicting the x cursor velocity. The predicted cursor velocity was compared with the original cursor velocity. As a measure of comparison, we calculated the correlation (r) between the two [58]. In preprocessing, we subtracted the mean of cursor velocity for better decoding performance.

In the 10-fold cross-validation, the data was repeatedly split into testing and training sets, such that testing set in each fold comprised a different 10% of the total dataset. We independently trained the model three times for each condition and determined the correlation between the original x cursor velocity and the prediction of the model. To analyze the statistical significance, we performed Fisher Z transformation on the correlation coefficient and conducted a paired t-test on the transformed values comparing the jittered spike and fixed spike conditions to determine if they yielded statistically different results.

**Data availability**

The data that support the graphs, figures, and images in this paper and other findings in this study are available in the Supplementary Information and from the corresponding author upon reasonable request.

**Code availability**

Custom-developed MATLAB code for RF simulation and demodulation for the proposed ASBIT protocol are available in the Supplementary Information from the corresponding authors upon reasonable request.

**References**

[1] Lee, J. *et al.* Neural recording and stimulation using wireless networks of microimplants. *Nature Electronics* **4**, 604–614 (2021).

[2] Buettner, M. *et al.* RFID sensor networks with the intel WISP. *In Proceedings of the 6th ACM conference on Embedded network sensor systems*, 393–394 (2008).

[3] Kang, Y.-S., Park, I.-H., Rhee, J. & Lee, Y.-H. MongoDB-based repository design for IoT-generated RFID/sensor big data. *IEEE Sensors Journal* **16**, 485–497 (2015).

[4] Vogt, H. Efficient object identification with passive RFID tags. *In International Conference on Pervasive Computing*, 98–113 (Springer, 2002).

[5] Lanzolla, A. & Spadavecchia, M. Wireless sensor networks for environmental monitoring. *Sensors* **21**, 1172 (2021).

[6] Darwish, A. & Hassanien, A.E. Wearable and implantable wireless sensor network solutions for healthcare monitoring. *Sensors* **11**, 5561–5595 (2011).

[7] Da Xu, L., He, W. & Li, S. Internet of things in industries: A survey. *IEEE Transactions on industrial informatics* **10**, 2233–2243 (2014).

[8] Li, S., Xu, L. D. & Zhao, S. The internet of things: a survey. *Information systems frontiers* **17**, 243–259 (2015).

[9] Mainetti, L., Patrono, L. & Vilei, A. Evolution of wireless sensor networks towards the internet of things: A survey. *In SoftCOM 2011, 19th international conference on software, telecommunications and computer networks*, 1–6 (IEEE, 2011).

[10] Liu, S.-C. & Delbruck, T. Neuromorphic sensory systems. *Current opinion in neurobiology* **20**, 288-295 (2010).

[11] Indiveri, G. & Douglas, R. Neuromorphic vision sensors. *Science* **288**, 1189-1190 (2000).

[12] Vanarse, A., Osseiran, A. & Rassau, A. A review of current neuromorphic approaches for vision, auditory, and olfactory sensors. *Frontiers in neuroscience* **10**, 115 (2016).

[13] Polyanskiy, Y. A perspective on massive random-access. *In 2017 IEEE International Symposium on Information Theory (ISIT)*, 2523-2527 (IEEE, 2017).

[14] Fengler, A., Jung, P., & Caire, G. SPARCs for unsourced random access. *IEEE Transactions on Information Theory*, **67**, 6894-6915 (2021).

[15] Amalladinne, V. K., Chamberland, J. F., & Narayanan, K. R. A coded compressed sensing scheme for unsourced multiple access. *IEEE Transactions on Information Theory*, **66**, 6509-6533 (2020).

[16] Klair, D. K., Chin, K.-W. & Raad, R. A survey and tutorial of RFID anti-collision protocols. *IEEE Communications surveys & tutorials* **12**, 400–421 (2010).

[17] Liva, G. Graph-based analysis and optimization of contention resolution diversity slotted ALOHA. *IEEE Transactions on Communications* **59**, 477–487 (2010).

[18] Chen, J., Skatchkovsky, N. & Simeone, O. Neuromorphic Integrated Sensing and Communications. *IEEE Wireless Communications Letters* (2022).

[19] Chen, J., Skatchkovsky, N. & Simeone, O. Neuromorphic wireless cognition: event-driven semantic communications for remote inference. *IEEE Transactions on Cognitive Communications and Networking* (2023).

[20] Culurciello, E., Etienne-Cummings, R. & Boahen, K. Arbitrated address event representation digital image sensor. In *2001 IEEE International Solid-State Circuits Conference. Digest of Technical Papers. ISSCC (Cat. No. 01CH37177)*, 92-93 (IEEE, 2001).

[21] Mutti, C. & Floerkemeier, C. CDMA-based RFID systems in dense scenarios: Concepts and challenges. In *2008 IEEE International Conference on RFID*, 215–222 (IEEE, 2008).

[22] Yang, Q., Wang, H.-M., Zheng, T.-X., Han, Z. & Lee, M. H. Wireless powered asynchronous backscatter networks with sporadic short packets: Performance analysis and optimization. *IEEE Internet of Things Journal* **5**, 984–997 (2018).

[23] Costello, J. T. *et al*. A low-power communication scheme for wireless, 1000 channel brain–machine interfaces. *Journal of Neural Engineering* **19**, 036037 (2022).

[24] He, Y. *et al*. An Implantable Neuromorphic Sensing System Featuring Near-Sensor Computation and Send-on-Delta Transmission for Wireless Neural Sensing of Peripheral Nerves. *IEEE Journal of Solid-State Circuits* **57**, 3058-3070 (2022).

[25] Kiani, M. & Ghovanloo, M. An RFID-based closed-loop wireless power transmission system for biomedical applications. *IEEE Transactions on Circuits and Systems II: Express Briefs* **57**, 260–264 (2010).

[26] Ghovanloo, M. & Atluri, S. An integrated full-wave CMOS rectifier with built-in back telemetry for RFID and implantable biomedical applications. *IEEE Transactions on Circuits and Systems I: Regular Papers* **55**, 3328–3334 (2008).

[27] Occhiuzzi, C., Cippitelli, S. & Marrocco, G. Modeling, design and experimentation of wearable RFID sensor tag. *IEEE Transactions on Antennas and Propagation* **58**, 2490–2498 (2010).

[28] Marrocco, G. RFID antennas for the UHF remote monitoring of human subjects. *IEEE Transactions on Antennas and Propagation* **55**, 1862–1870 (2007).

[29] Xiao, Z. *et al*. An implantable RFID sensor tag toward continuous glucose monitoring. *IEEE journal of biomedical and health informatics* **19**, 910–919 (2015).

[30] Rose, D. P. *et al*. Adhesive RFID sensor patch for monitoring of sweat electrolytes. *IEEE Transactions on Biomedical Engineering* **62**, 1457–1465 (2014).

[31] Niu, S. *et al*. A wireless body area sensor network based on stretchable passive tags. *Nature Electronics* **2**, 361–368 (2019).

[32] Yeon, P., Bakir, M. S. & Ghovanloo, M. Towards a 1.1 mm$^2$ free-floating wireless implantable neural recording soc. In *2018 IEEE Custom Integrated Circuits Conference (CICC)*, 1–4 (IEEE, 2018).

[33] Bandodkar, A. J. *et al*. Battery-free, skin-interfaced microfluidic/electronic systems for simultaneous electrochemical, colorimetric, and volumetric analysis of sweat. *Science advances* **5**, eaav3294 (2019).

[34] Lee, J. *et al*. An implantable wireless network of distributed microscale sensors for neural applications. In *2019 9th International IEEE/EMBS Conference on Neural Engineering (NER)*, 871–874 (IEEE, 2019).

[35] Leung, V. W. *et al*. Distributed microscale brain implants with wireless power transfer and Mbps bi-directional networked communications. In *2019 IEEE Custom Integrated Circuits Conference (CICC)*, 1–4 (IEEE, 2019).

[36] Nurmikko, A. Challenges for large-scale cortical interfaces. *Neuron* **108**, 259–269 (2020).

[37] Nurmikko, A. V. *et al*. Listening to brain microcircuits for interfacing with external world—progress in wireless implantable microelectronic neuroengineering devices. *Proceedings of the IEEE* **98**, 375–388 (2010).

[38] Hochberg, L. R. *et al*. Neuronal ensemble control of prosthetic devices by a human with tetraplegia. *Nature* **442**, 164–171 (2006).

[39] Seo, D., Carmena, J. M., Rabaey, J. M., Alon, E. & Maharbiz, M. M. Neural dust: An ultrasonic, low power solution for chronic brain-machine interfaces. *arXiv preprint arXiv:1307.2196* (2013).

[40] Olshausen, B. A. & Field, D. J. Sparse coding of sensory inputs. *Current opinion in neurobiology* **14**, 481–487 (2004).

[41] Foldiak, P. Sparse coding in the primate cortex. *The handbook of brain theory and neural networks* (2003).

[42] Dinan, E. H. & Jabbari, B. Spreading codes for direct sequence CDMA and wideband CDMA cellular networks. *IEEE communications magazine* **36**, 48–54 (1998).

[43] Man, A. S. *et al*. Design and implementation of a low-power baseband-system for RFID tag. In *2007 IEEE International Symposium on Circuits and Systems*, 1585–1588 (IEEE, 2007).

[44] Baghaei-Nejad, M. *et al*. A remote-powered RFID tag with 10Mb/s UWB uplink and -18.5 dBm sensitivity UHF downlink in 0.18 µm CMOS. In *2009 IEEE International Solid-State Circuits*

*Conference-Digest of Technical Papers*, 198–199 (IEEE, 2009).

[45] Hwang, Y.-S. & Lin, H.-C. A new CMOS analog front end for RFID tags. *IEEE Transactions on Industrial Electronics* **56**, 2299–2307 (2009).

[46] Yin, J. *et al*. A system-on-chip EPC gen-2 passive UHF RFID tag with embedded temperature sensor. *IEEE Journal of Solid-State Circuits* **45**, 2404–2420 (2010).

[47] Orchard, Garrick, *et al*. Efficient neuromorphic signal processing with loihi 2. In *2021 IEEE Workshop on Signal Processing Systems (SiPS)*, 254–259 (IEEE, 2021).

[48] Heelan, C. *et al*. Decoding speech from spike-based neural population recordings in secondary auditory cortex of non-human primates. *Communications biology* **2**, 1–12 (2019).

[49] Heelan, C., Nurmikko, A. V. & Truccolo, W. FPGA implementation of deep-learning recurrent neural networks with sub-millisecond real-time latency for BCI-decoding of large-scale neural sensors (104 nodes). In *2018 40th Annual International Conference of the IEEE Engineering in Medicine and Biology Society (EMBC)*, 1070–1073 (IEEE, 2018).

[50] Fagg, A. H., Ojakangas, G. W., Miller, L. E. & Hatsopoulos, N. G. Kinetic trajectory decoding using motor cortical ensembles. *IEEE Transactions on Neural Systems and Rehabilitation Engineering* **17**, 487–496 (2009).

[51] Benjamin, A. S. *et al*. Modern machine learning as a benchmark for fitting neural responses. *Frontiers in computational neuroscience* **56** (2018).

[52] Ghosh-Dastidar, S. & Adeli, H. Spiking neural networks. *International journal of neural systems* **19**, 295-308 (2009).

[53] Orchard, G., Jayawant, A., Cohen, G. K. & Thakor, N. Converting static image datasets to spiking neuromorphic datasets using saccades. *Frontiers in neuroscience* **9**, 437 (2015).

[54] Dethier, J., Nuyujukian, P., Ryu, S. I., Shenoy, K. V. & Boahen, K. Design and validation of a real-time spiking-neural-network decoder for brain–machine interfaces. *Journal of neural engineering* **10**, 036008 (2013).

[55] Kumarasinghe, K., Kasabov, N & Taylor, D. Brain-inspired spiking neural networks for decoding and understanding muscle activity and kinematics from electroencephalography signals during hand movements. *Scientific reports* **11**, 2486 (2021).

[56] Burns, B. D. & Webb, A. The spontaneous activity of neurones in the cat's cerebral cortex. *Proceedings of the Royal Society of London. Series B. Biological Sciences* **194**, 211–223 (1976)

[57] Mizuseki, K. & Buzsáki, G. Preconfigured, skewed distribution of firing rates in the hippocampus and entorhinal cortex. *Cell reports* **4**, 1010–1021 (2013).

[58] Glaser, J. I. *et al*. Machine learning for neural decoding. *Eneuro* **7** (2020).

[59] LAVA-DL. Lava-nc/lava-dl. GitHub. Retrieved March 28, 2023, from https://github.com/lava-nc/lava-dl

[60] Wong, T.-T. & Yeh, P.-Y. Reliable accuracy estimates from k-fold cross validation. *IEEE Transactions*

*on Knowledge and Data Engineering* **32**, 1586-1594 (2019).

**ACKNOWLEDGEMENTS**

We acknowledge N. Fathy, J. Huang, S. Li, S. Yu, L. Cui, S. Alluri, and M. Lokhandwala at UCSD for their previous work on ASIC backbone design. We also thank Y.-K. Song and D. Durfee for their insights into ASIC and hardware design and Tim Shea and Mike Davies at Intel Labs for pioneering work with SNN technologies. We have greatly benefited from the insight of our colleagues across multiple fields from microelectronics to brain sciences and clinical neurology: P. Asbeck, J. Donoghue, B. Dutta, J. Groe, L. Hochberg, T. Sejnowsky, K. Shenoy, and S. Cash, among others. This research was supported by private gifts.

# Supplementary Information

## An Asynchronous Wireless Network for Capturing Event-Driven Data from Large Populations Autonomous Sensors

Jihun Lee[1*], Ah-Hyoung Lee[1*], Vincent Leung[2], Farah Laiwalla[1], Miguel Angel Lopez-Gordo[3], Lawrence Larson[1], and Arto Nurmikko[1,4]

[1] School of Engineering, Brown University, Providence, RI, USA
[2] Electrical and Computer Engineering, Baylor University, Waco, TX, USA
[3] Department of Signal Theory, Telematics and Communications, University of Granada, Granada, Spain
[4] Carney Institute for Brain Science, Brown University, Providence, RI, USA

**Supplementary Note 1. The choice of Gold code and matched filter synthesis**

Gold codes are one type of pseudo-random (PN) sequence that are commonly used in spread-spectrum communication. A set of Gold codes provides quasi-orthogonality, which allows multiple users to share the same frequency band without being affected by mutual interference. The use of Gold codes helps to improve signal quality in noisy environments in general by allowing the receiver to filter out unwanted noise and interference. In the context of the ASBIT protocol, such noise and interference would be mainly due to background signals transmitted from other nodes. Furthermore, compared to other PN generators such as the Kasami code generator, the circuit implementation of the Gold code generator is simple, as shown in Supplementary Fig. 1a, resulting in a low-power, small footprint ASIC design [1]. However, the Gold code may not be the optimal choice for all circumstances, and other pseudorandom (PN) sequences could also be utilized in the ASBIT protocol of this paper.

In the ASBIT protocol, matched filters are important for event recovery by RF demodulation. To synthesize the matched filters, the pre-discovered digital Gold code at 10 Mbps is first converted to an analog waveform which matches with the backscattering waveforms from the specific target chip collected at the RF receiver. Supplementary Fig. 3 demonstrates the synthesis process of generating backscattering signals from the digital Gold code on RF microsensors. The method section ("Modeling of backscattering signals from RF microsensors for matched filter design") provides relevant theory and equations for the process. Briefly, the gold code sequence is first encoded into binary phase shift keying (BPSK) as a digital waveform using a 30 MHz digital clock. It is then transmitted through backscattering by toggling a capacitor in the transmission circuit on the microchip. The external RF receiver, a software-defined radio (SDR) in our case, collects the signal and down-converts the backscattered sideband (~945 MHz) to DC. This entire process can be modeled simply by down-converting the BPSK-encoded gold code signal from 30 MHz to DC which results in an estimated analog waveform of the Gold code. In the event recovery step of the ASBIT protocol, we used this analog Gold code waveform to generate a specific matched filter. Supplementary Figs. 1b and 1c show photos of the relationship and the size of the Gold code finite state machine (FSM) in relation to the overall wireless chip.

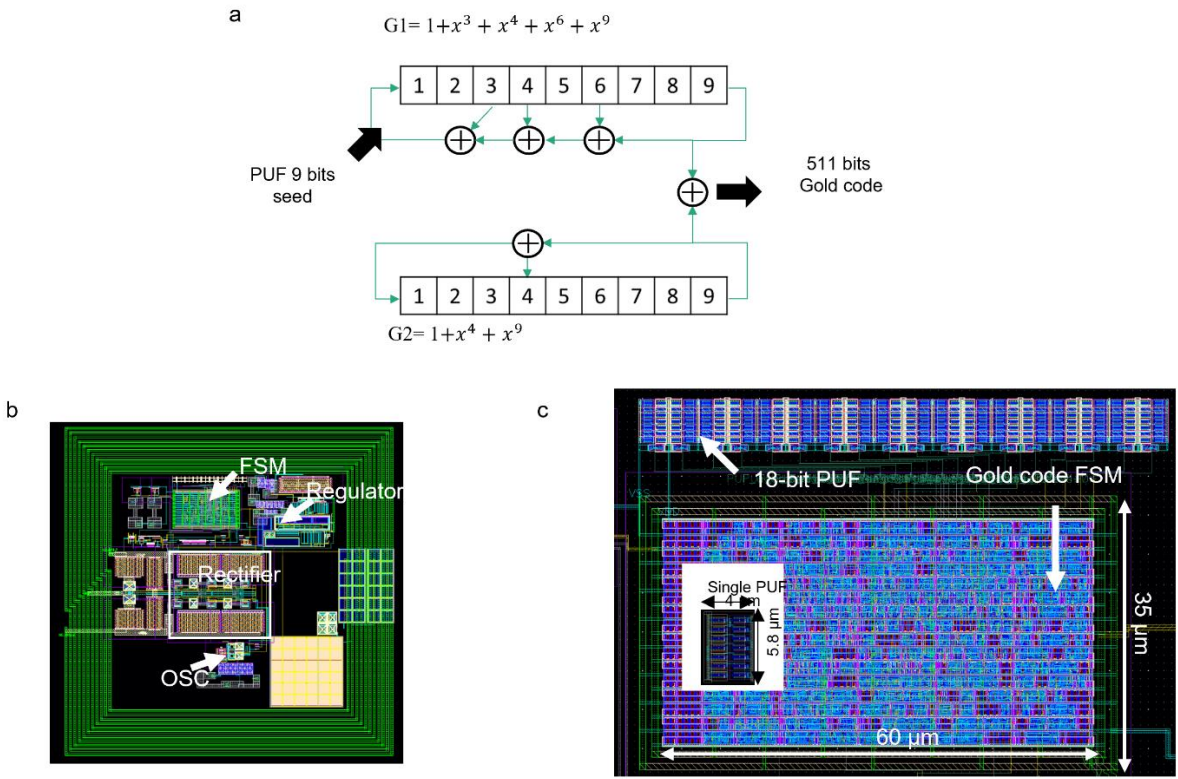

**Supplementary Figure 1.** a) Schematic of a 511-bit Gold code generator, where a linear-feedback shift register is combined with a physically unclonable function to generate the Gold code sequence. b) The layout of the prototype wireless RFID microchip where the power harvesting rectifier and the perimeter circling on-chip coil occupy most of the circuit area whereas the footprint of the finite state machine (FSM) is considerably smaller. c) Magnified view of the Gold code FSM showing its footprint of 35 µm × 60 µm, together with the physically unclonable function (PUF) seed circuit which has a unit size of 4 µm × 5.8 µm. The PUF provides a random number unique to each microchip in order for the chip to synthesize its unique gold code. To generate a Gold code, we utilized 13 bits of the PUF as a seed sequence. This allowed us in principle to synthesize a total of 8191 bits of Gold code. However, to reduce mutual interference during communication, only a portion of the Gold code consisting of 511 bits was actually transmitted. By using a 13-bit seed sequence instead of a 9-bit seed sequence, we were able to minimize the likelihood of PUF-induced collisions (i.e., the very same PUF sequence was present in several chips).

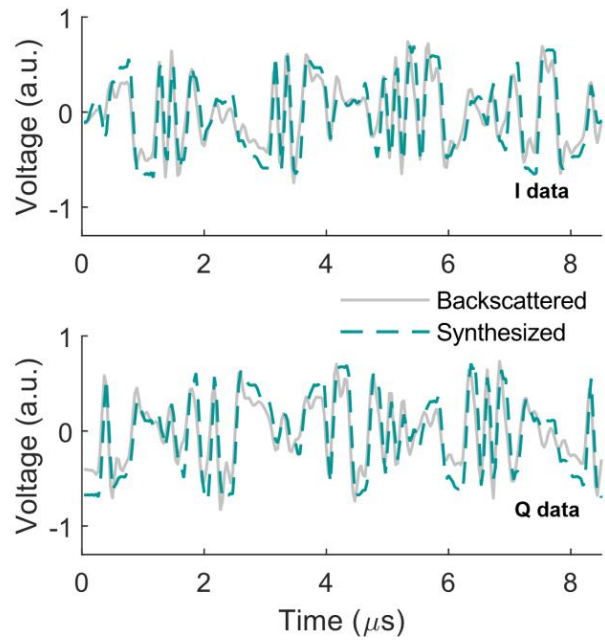

**Supplementary Figure 2.** Example of experimental I/Q data measured from the wireless chip (gray trace) compared with that synthesized by a matched filter as generated by converting a digital bit stream into analog I/Q waveform in the ASBIT network simulation model (green trace).

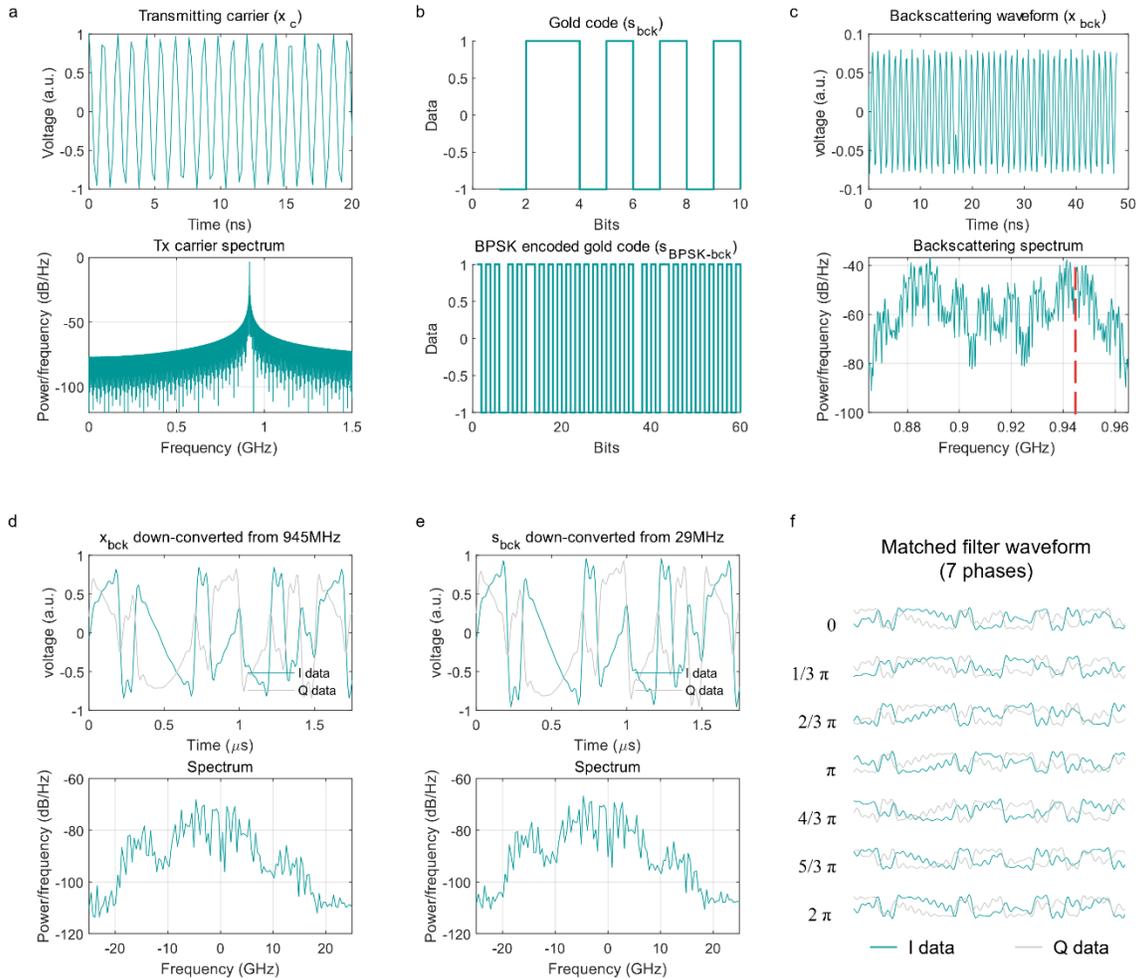

**Supplementary Figure 3.** A synthesis process to generate matched filter waveforms according to equations (1)-(6) in Methods ("Modeling of backscattering signals from RFID microsensors for matched filter design"). a) Downlink 915 MHz carrier wave ($x_c$) in Main Eq (1) (time resolution 200 ns) and its spectrum. b) Gold code sequence ($s_{bck}$) encoded by binary phase shift keying (BPSK) in the digital waveform $s_{BPSK-bck}$. c) Backscattered signals ($x_{bck}$) generated by modulating $x_c$ by $s_{BPSK-bck}$, and the corresponding spectrum of $x_{bck}$. The red dotted vertical line shows the center frequency of the target side band (946 MHz; i.e. 915 MHz + clock of the target node at 31 MHz); d) I and Q data of $x_{bck}$ waveforms down-converted from 945 MHz to DC displaying a 1 MHz residual clock frequency, i.e. the particular node clock frequency subtracted by the nominal clock frequency of 30 MHz; its finite spectrum reflecting the bandwidth of the bit sequence (10 Mbps). e) Waveforms of I and Q data and their spectra obtained by down-converting $s_{bck}$ from 29 MHz with a residual clock frequency of 1 MHz. According to Main Eq. (5), $x_{bck}$ waveforms down-converted from 945 MHz to DC should be equivalent to that of $s_{bck}$ down-converted from 29 MHz to DC. f) Examples of matched filter waveforms showing phase dependent I and Q data with the residual clock of 3 MHz, which shows that the waveform is determined by the phase delay and the sampling phase (See Methods for detailed description).

**Supplementary Note 2. Analysis of Received Signal Strength Indicator (RSSI), signal-to-noise (SNR), and efficiency**

In the ASBIT backscattering method, the wireless transfer efficiency is crucial for the device's operation for both energy harvesting from the external transceiver (Tx) downlink while using it for backscattered communication (Supplementary Fig. 4a). The Tx transmits power ($P_{tx}$), which is captured by an on-chip coil (Rx) as $P_{tx} + \eta$ (defined as the wireless efficiency at 915 MHz), where η is determined by the coupling factor between Tx and Rx coils (in case of near-field) or antenna gain (in case of far-field) plus the path loss. The received energy is then converted to a backscattering signal at a conversion efficiency $\eta_c$ by the device circuits in the sensor microchip and transmitted back to the transceiver (Tx). The backscattering signal is then subject to wireless efficiency (η + c, where c is additional path loss at 945 MHz).

Thus, the level of the Received Signal Strength Indicator (RSSI) in the system can be expressed as:

$$RSSI\ (dBm) = P_{tx}\ (dBm) + 2 \times \eta + c + \eta_c \qquad \text{(Supplementary Eq. 1)}$$

Using this equation, we measured the change in RSSI with varying Tx power (Supplementary Fig. 4b). The plot on the left shows the chip turning on at a Tx power of 13 dBm to achieve the target clock frequency. The plot on the right further shows how, when Tx power increased, the RSSI does not increase linearly, indicating that $\eta_c$ is not constant and can vary with the Tx power in our backscattering modulation circuit. This design of circuit configuration alleviates the near-far problem, as the node with higher energy does not generate backscattering signals simply proportional to its power level.

Since the chips operate in fully wireless mode, we cannot precisely determine the conversion efficiency. Still, according to Supplementary Eq. 1, we can estimate the relationship between conversion efficiency and wireless transfer efficiency. Based on a combination of ASIC circuit and RF simulations, we estimated the threshold RF level for a microchip to be -16 dBm, 25.11 µW [2]. In that case, η can be solved as -29 dB so that the efficiency $\eta_c$ + c can be in the range of -24 dB.

From another perspective and according to the definition of Signal-to-Noise Ratio (SNR), RSSI can be also solved as

$$SNR = RSSI\ (dBm) - Noise\ floor\ (dBm) \qquad \text{(Supplementary Eq. 2)}$$

Therefore, the wireless transfer efficiency (η) and its relationship to SNR can be determined using pre-discovered values of $P_{tx}$, $\eta_c$ + c, and noise floor (dBm) for the specific case. In our microchips and from their RF interface, $\eta_c$ + c was measured as -24 dB and $P_{tx}$ of 24 dBm was chosen, since this is at the regulatory limit for the Specific Absorption Rate (SAR), 10 Wkg$^{-1}$ SAR averaged over 10 g of the tissue [3]. Also, since we do not use any additional means to control the background noise (such as a Faraday cage), the noise floor of the system, measured by a software-defined ratio, was -75.75 dBm. At this noise level, the relationship between SNR and corresponding wireless transfer efficiency can be solved using Supplementary Eqs. 1 and 2. For example, with an SNR of 3 dB, the estimated RSSI is -72.75 dBm, and the predicted wireless power transfer efficiency is -36.375 dB. Therefore, SNR values used in plots Fig. 3e and Fig. 4c also provide an indirect indication of the wireless efficiency of the system. We note that the relationship between SNR and the wireless transfer efficiency can vary in other wireless systems based on factors such as coil size, conversion efficiency, and noise floor. Therefore, the analysis presented here is accurate only for our specific experimental setup.

We characterized wireless power harvesting in the fabricated chips using the 3-coil system shown in Fig. 2c. Supplementary Fig. 4c illustrates the measured and simulated efficiency based on the chip's location within the relay coil. We determined the efficiency of the wireless link by subtracting the required on-chip energy (found to be -16 dBm in the circuit simulation) from the minimum Tx power needed to activate the chip. The measurements indicated an efficiency range of -35.45 dB to -26.95 dB, with the average measured efficiency 4.1 dB lower than the simulated efficiency on average. This discrepancy could be attributed to an impedance mismatch between the on-chip coil and circuits. Note that we have also made improvements in our current system by optimizing the on-chip coil design, resulting in a lower loss of only 4.1 dB compared to the 7 dB loss in our previously reported case [2].

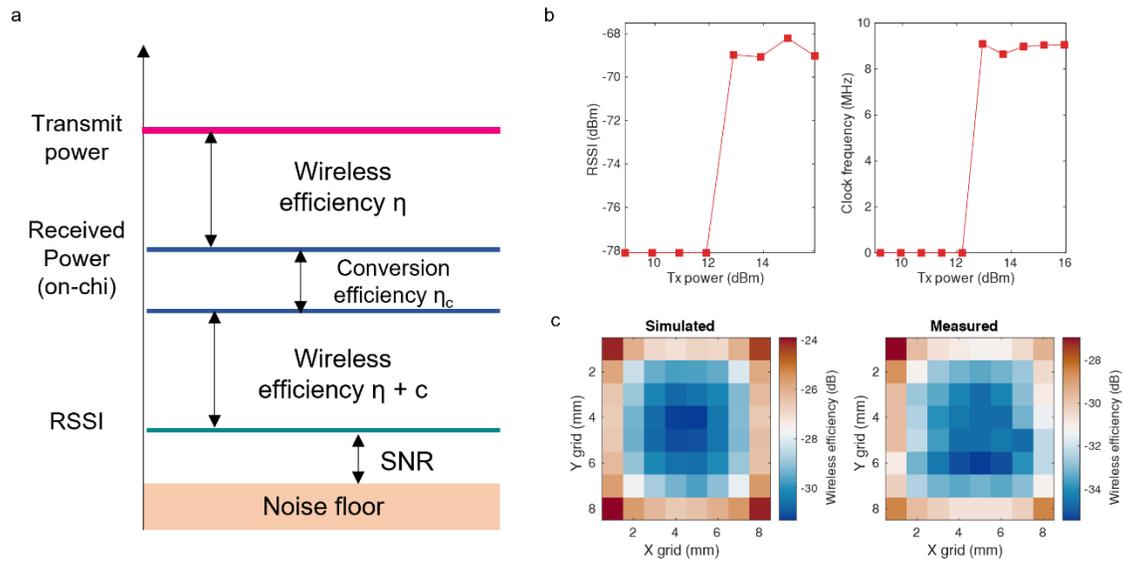

**Supplementary Figure 4.** Relationship between wireless transfer efficiency, Received Signal Strength Indicator (RSSI) and signal-to-noise ratio (SNR). a) Schematic diagram illustrating the levels of RSSI and SNR in our ASBIT backscattering scheme. The SNR level is influenced by various factors including wireless transfer efficiency ($\eta$ @915 MHz), backscattering efficiency ($\eta + c$, @945 MHz), conversion efficiency ($\eta_c$), and the noise floor. b) Dependence of RSSI of the backscattering signals and clock frequency of the microchip on the Tx power; c) Comparison between simulated and measured wireless power transfer efficiency at 64 different locations within the relay coil perimeter, indicating a 4.1 dB difference due to possible impedance mismatch and circuit process variance.

**Supplementary Note 3. Comparison between an ASIC with free running on-chip oscillator and a clock frequency divider approach**

First, we assessed the consistency of the chips individually, ensuring their ability to generate the same Gold code pattern over a finite period for the ASBIT communication. All 78 chips with the free-running oscillator consistently produced backscattering signals every 20 ms. The left plot in Supplementary Figure 5a shows that, among these, 47 chips generated identical Gold codes across all 1,000 packets, and 18 chips transmitted the same code more than 90% of the time. We also tested the Gold code repeatability of 65 chips with the frequency divider replacing the on-chip oscillator as shown in the right plot of Supplementary Fig. 5a. Out of these, 40 chips backscattered the same Gold code over 1,000 times. In the experiments described in the Main text, we selected a target group of chips that consistently generated the same Gold code. All remaining chips were used to generate background signals for simulating a large-scale network. The instability of the PUF circuit, the seed for the Gold code, caused the Gold code synthesizer to generate variant codes. Such instability is well known in PUF which comes from measurement noise and environmental fluctuations [4]. Implementing a PUF error correction circuit, which samples PUF sequences repeatedly and finds the most frequent bits, can improve the yield since the PUF instability was temporary most of the time in this and our previous study [2].

Supplementary Fig. 5a also shows the variation in clock frequencies among the chips with the oscillator and the frequency divider. The clock frequency ranges from 10 to 11.17 MHz in our on-chip relaxation oscillator design, commonly chosen for its low power and area, that incorporates a capacitance subject to foundry process variance. We addressed this clock variance by designing a set of matched filters specific to each chip that could reliably detect the Gold code waveform. In the frequency divider-based ASBIT microchip, on the other hand, most chips generated a 9.53 MHz clock from a down-converted 28.59 MHz signal. Therefore, the chip-to-chip clock variance was negligible, except for a few outliers, which facilitated the demodulation step. Also, as shown in Supplementary Fig. 5b, the clock drift in the divider-based system was significantly lower compared to the oscillator-based system, leading to a consistent backscattered Gold code waveform. This accelerated the demodulation process as only three matched filters were required to account for sampling phase variance.

The cost of using the frequency divider approach requires more incoming RF energy compared to the oscillator-based system, leading to a tradeoff. As shown in Supplementary Fig. 5c, the average power required to turn on the divider-based microchip is approximately 5 dBm higher than that of the oscillator-based chips. This higher power requirement may be attributed to the higher input signal needed to cross the threshold voltage of the input differential amplifier, which may also explain the observed variance among chips. Nonetheless, we observed that some chips with dividers were able to operate with only a 2 dB increase in Tx power compared to those with on-chip free oscillators, indicating a potential for further improvement.

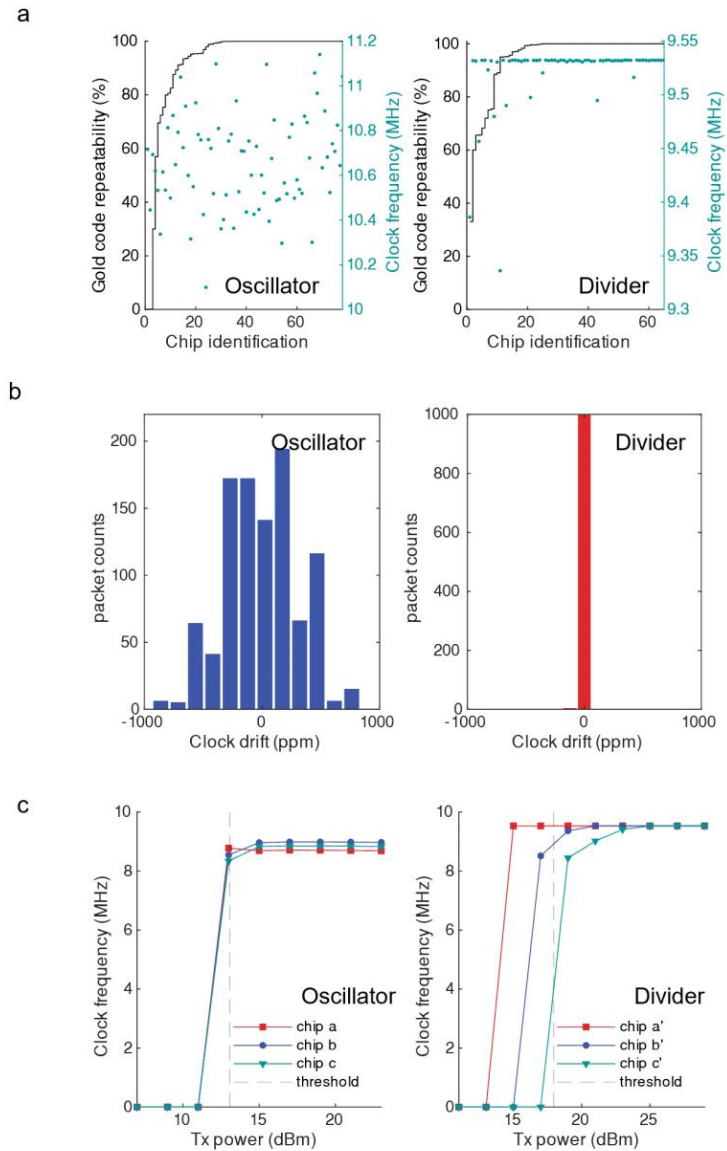

**Supplementary Figure 5. Gold code generation and clock frequency stability in the experimentally measured ASBIT microchips. a)** Left: Gold code packet repeatability and clock frequency variance among the chips of 78 microchips with free running oscillator clocks tested with 1,000 packets per chip. Right: Gold code repeatability and clock frequency variance of 65 microchips using the on-chip frequency divider tested with 1,000 packets each, showing the nominal clock frequency of 9.54 MHz in the majority of chips except for a few outliers. **b)** Histogram of the clock frequency drift of 1,000 packets from the oscillator and frequency divider, respectively. **c)** Clock frequency as a function of Tx power for three microchips randomly selected from a population of chips with free-running oscillator and frequency divider circuits, respectively, with the vertical gray dash line indicating the average threshold level for all available chips in each case. Abbreviation: ppm: parts per million.

**Supplementary Note 4. Fundamental network capacity and coding gain in the wireless ASBIT network**

While we evaluated the scalability of the ASBIT network using experimental data collected from microfabricated chips, one can also estimate the ideal network capacity for the proposed network. In the simplified limit of equal backscattered amplitude $S$ from all nodes (i.e. ignoring the near-far problem), the SNR can be defined as follows, accounting for interference among all nodes [5-7]

$$SNR = S/[(N-1) \times S + \eta] \quad \text{(Supplementary Eq. 3)}$$

where N is the number of nodes in the network and $\eta$ is the background noise from spurious interference plus thermal noise within the total spread bandwidth $W$. The ratio of energy per bit to noise power spectral density ($E_b/N_0$) can be expressed as

$$E_b/N_0 = \frac{W/R}{(N-1)+(\eta/S)} \quad \text{(Supplementary Eq. 4)}$$

while $R$ denotes the bit data rate before spreading. Then, the network capacity in terms of the number of supported nodes becomes

$$N = 1 + \frac{W/R}{E_b/N_0} - \frac{\eta}{S} \quad \text{(Supplementary Eq. 5)}$$

Here, $W/R$ is generally referred to as the CDMA processing gain, or coding gain, $L_c$. Given that $W \approx 1/T_c$ ($T_c$ is the 'chip' data duration, the 'chip' defined as each bit in the Gold code) for a given predefined bit rate $R$, this equation implies that the maximum allowable number of nodes increases as the chip duration becomes shorter and the Gold code gets longer, which demands a wider bandwidth $W$. In our case, the bandwidth $W$ is limited to 10 MHz by the present ASIC design and implementation constraints. Since the bit rate $R$ is inversely proportional to the length of the Gold code, $L_c$ and thus the network capacity is simply proportional to the length of the Gold code. Assuming an application acceptable event error rate (EER) of 10[-3] for a path loss is 40 dB, and imposing a requirement for $E_b/N_0$ of 7 dB for 511-bit BPSK Gold code, we obtain that a single CDMA channel can accommodate 169 nodes in the ASBIT protocol with a thermal noise density of -174 dBm/Hz. Importantly, however, under the assumption of sparsity in the ASBIT scheme with only some 5% of band utilization per node (50 Hz event rate), the network capacity of the ASBIT protocol for the case of a 511-bit Gold code is estimated to be 3,380 nodes. In this ideal limit, one has ignored the impact of e.g. clock variance across nodes, clock drift at each node, and the amplitude variance (near-far problem) which reduces the limits the network capacity. Note, however, that in our paper, driven by experimental data on actual chips, we have taken into account a number of non-ideal factors to evaluate the practical network capacity of a functional ASBIT protocol while applying specialized techniques such as demodulation through discrete timing.

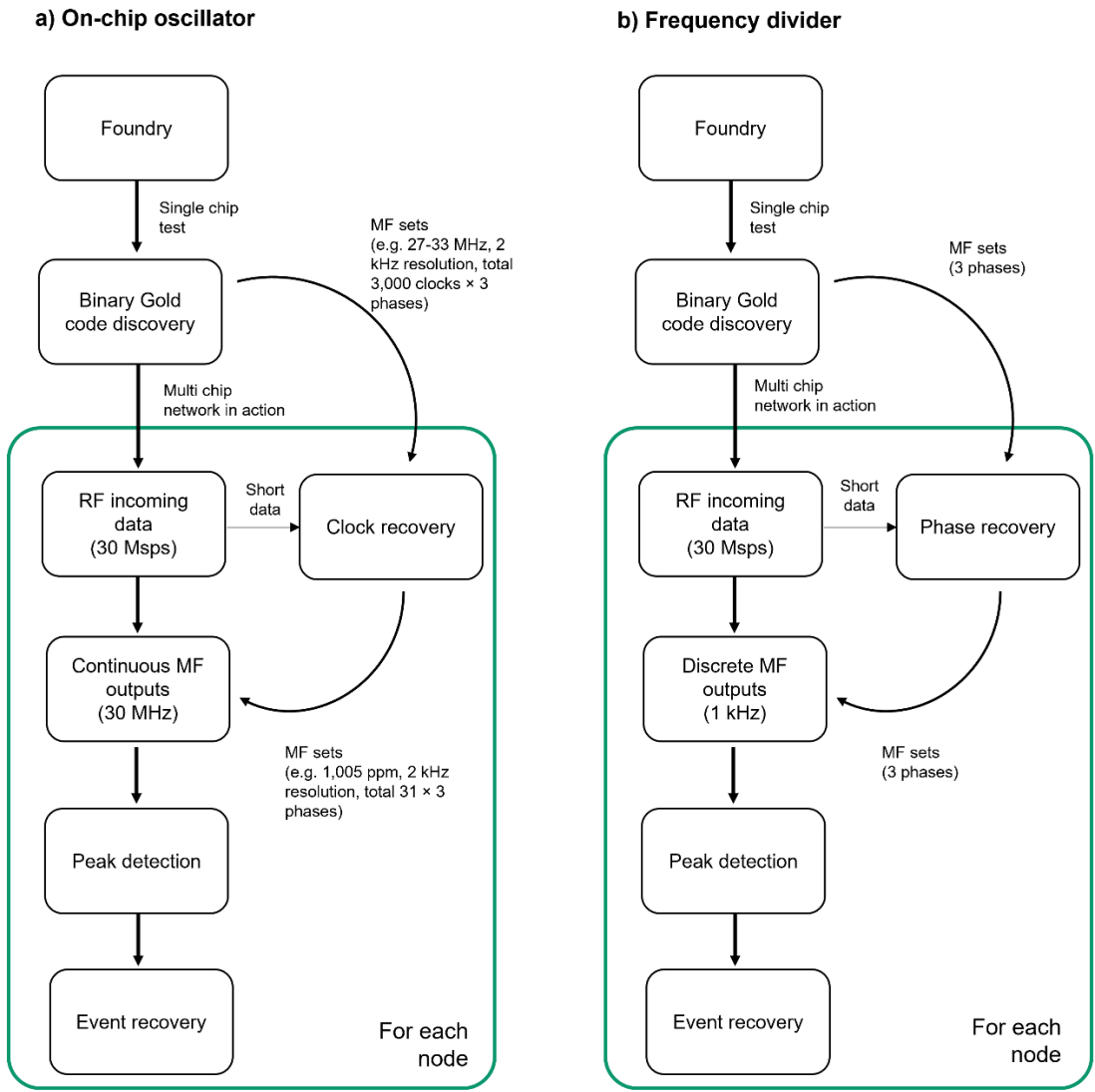

**Supplementary Figure 6.** Flow chart of the event recovery process at the receiver, i.e. demodulation of data from ensembles of RF microsensors encoded according to the ASBIT protocol; a) a free running oscillator as the on-chip system clock; b) Case of baseband carrier derived on-chip frequency divider for clock generation. Abbreviation: MF: matched filter.

**Supplementary Note 5. The comparison between the ASBIT and other communication methods used in sensor devices specific to neural implants**

The ASBIT protocol is distinct from other conventional communication methods, whether proposed or implemented for specific applications, such as for wireless neural implants, because it does not need to allocate specific resources like timeslots and frequency bands to nodes. Here, we first summarize the key points about the ASBIT protocol and then show examples from the neural implant space. We note that as the ASBIT protocol targets wireless RFID-type smart sensor networks, an emerging field still under exploration. In research on various random-access protocols on RFID tags (passive, no sensing capability), a sophisticated anti-collision protocol using time-division multiple access (TDMA) is proposed as one avenue [8-9] or other deterministic tree-search-based methods [10, 11] or probabilistic time-slotted methods [12-14]. Whether deterministic or probabilistic, their scalability is still limited due to penalties from time scheduling or packet collisions and the maximum number of sensor nodes remains well below fundamental limits. In our own recent work with RFID-type neural microsensors as brain implants, we proposed a TDMA method, leveraging a call-and-response type of bidirectional communication [2] yet found that the number of sensors would be limited to ≈770 incurring a system latency of 100 ms due to the spectral penalty imposed by the downlink.

The aggregate network capacity of the ASBIT system with frequency dividers has been estimated to be 100,000 events per second, with a single node able to report any number of events below the upper limit. To put this in perspective, the sensor in the ASBIT system can be thought of as a device transmitting 1 kbps rate of data, assuming a bin size of 1 msec and an event rate of 50 Hz (5% ones, backscattering, and 95% zeros, silence). This yields an aggregate data rate equivalent to 2 Mbps, which can increase to 10 Mbps if the event rate drops to 10 Hz. With an RF bandwidth of 10 MHz and BPSK encoding, the spectral efficiency of the ASBIT system is already approaching the theoretical maximum of 1 bit/Hz. By contrast, our own previous TDMA method [2] had a spectral efficiency of 0.77 bit/Hz even allowing for a 100 msec system latency.

Turning to RF communication systems used or proposed in implants (Table S1), the ASBIT system uses a much higher system clock for higher data rates, which can lead to higher current consumption. However, we reduce power consumption by utilizing sparse backscattering for data communication and optimizing circuits to improve impedance matching. Our microchip operates with 25 µW, a number which would increase by 3.2 µW when including the sensor circuit for neural recording which was not included in our current prototype ASBIT communication chip.

A broadband neural recording device described in [15] has a data rate of 800 kbps, while an 8-channel field potential (LFP) recording system presented in [16] has a data rate of 205 kbps. Both systems generate backscattering for communication using impulse radio ultra-wideband (IR-UWB) or load shift keying (LSK), which allocate timeslots or frequency bands entirely to a single node. However, these systems have much larger current consumption than in our case. The 'neural dust' system in [17] uses only passive transistors for backscattering, which reduces power consumption but limits the signal quality. And, especially, the scalability of the system beyond a handful of sensors appears quite limited. The optically powered 'Mote' interface described in [18] achieves a low power consumption of less than 1 µW by utilizing optical pulse position modulation for data communication. However, its performance has only been evaluated in limited conditions that do not account for losses in fully assembled optical-electrical devices the monolithic microfabrication of which poses a major heterogeneous integration challenge. By contrast, our device principle is fully compatible with a monolithic system-on-chip structure, with all circuit blocks fully integrated into a single piece of silicon. Microfabrication related issues and associated losses/yield are therefore absent in our case result. When applying this technology to wireless neural interfaces, the only additional loss that needs to be accounted for is the transmission loss in biological tissue. These losses have been studied and are presented in Supplementary Fig. 9.

Table S1I Comparison with communication method in state-of-the-art microimplants

|  | [4] | [5] | [6] | [7] | Current work |
|---|---|---|---|---|---|
| Number of node(s) | 1 | 1 | 1 | Up to 1000 | Up to 8000 |
| Wireless Powering | 131 MHz RF 3-coil | 433 MHz RF 3-coil | 1.85 MHz Ultrasonic | Optical | 915 MHz RF 3-coil |
| External Tx power (mW) | - | - | 0.12 | - | Up to 316 |
| Telemetry | Backscattering IR-UWB | Backscattering LSK | Backscattering | Pulse Position Modulation | Backscattering BPSK (ASBIT) |
| Uplink data rate (Mbps) | 0.8 Mbps | 0.205 Mbps | 0.5 Mbps | 0.6 -1.3 kbps (aggregate 0.6-1.3 Mbps) | 1 kbps per node (Aggregate 1-10 Mbps)* |
| | | | | | |
| Technology | 350 nm CMOS+ discrete | 350 nm CMOS | Discrete | 180 nm CMOS | 65 nm CMOS |
| IC Area [mm$^2$] | 1.1 | 12.25 | 0.032 | 0.0297 | 0.42 |
| Energy harvester | Discrete coil | Discrete coil | Piezoelectric | Photovoltaic cell | On-chip Coil |
| Power supply (V) | 1.8 | 1.5 | - | 1.5 | 0.8 |
| Power Consumption (uW) | <300 | 92 | <1 | <1 | <30 |
| | | | | | |
| * Determined by the sparsity of the target signal | | | | | |

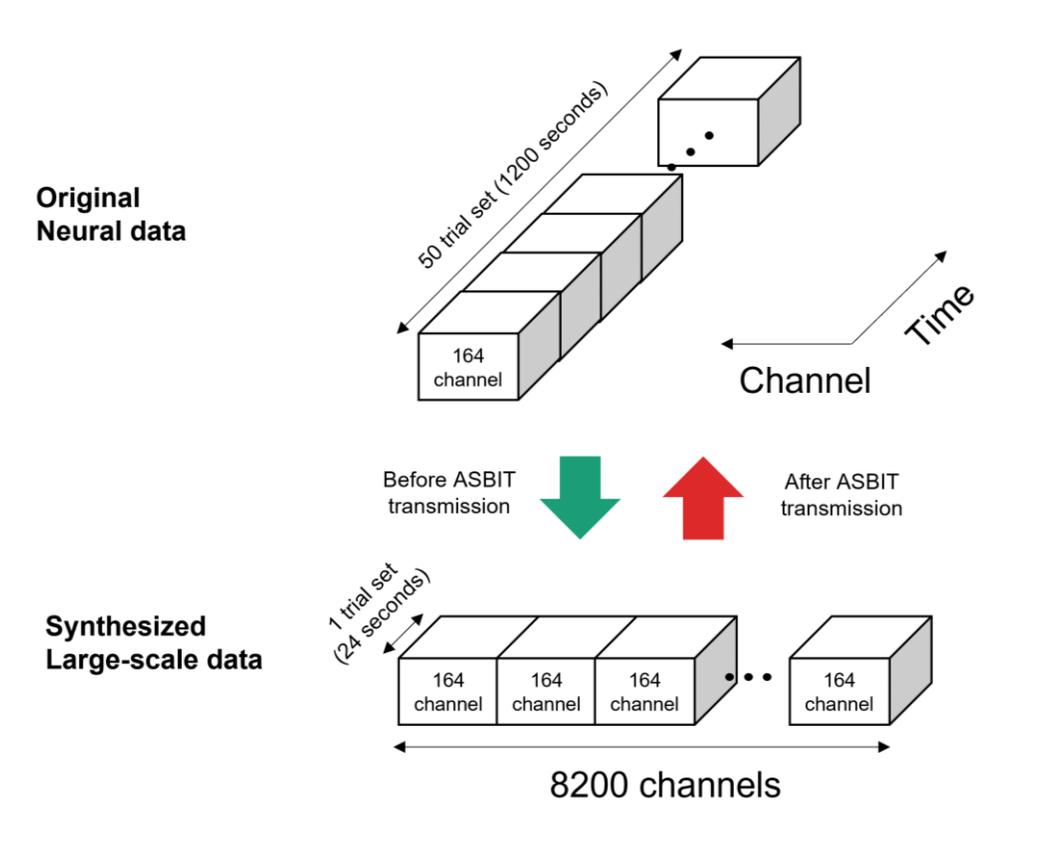

**Supplementary Figure 7.** The scheme for preprocessing open source multichannel primate electrophysiological neuronal spike data, recorded from the primate cortex by wired microelectrode arrays, here providing the input to an ensemble of wireless recording microchips using the ASBIT protocol. Available open source neural data was limited to only 164 spiking channels whereas the network capacity of the ASBIT protocol can reach up to several thousands. We increased the number of channels by transmitting neural data from 50 or 160 datasets simultaneously to generate an equivalent of 8,200 channels (164 neurons × 50 datasets) in the motor cortex (M1) and, 8,320 channels (52 neurons × 160 datasets) in the somatosensory cortex (S1). Following the ASBIT transmission step from the corresponding number of microsensors, we reconstituted and mapped the data back into the original number of channels for the use of this data to train a neural network model for brain-machine interface (BMI)-relevant neural decoding. All experimental neural data were obtained from [19, 20].

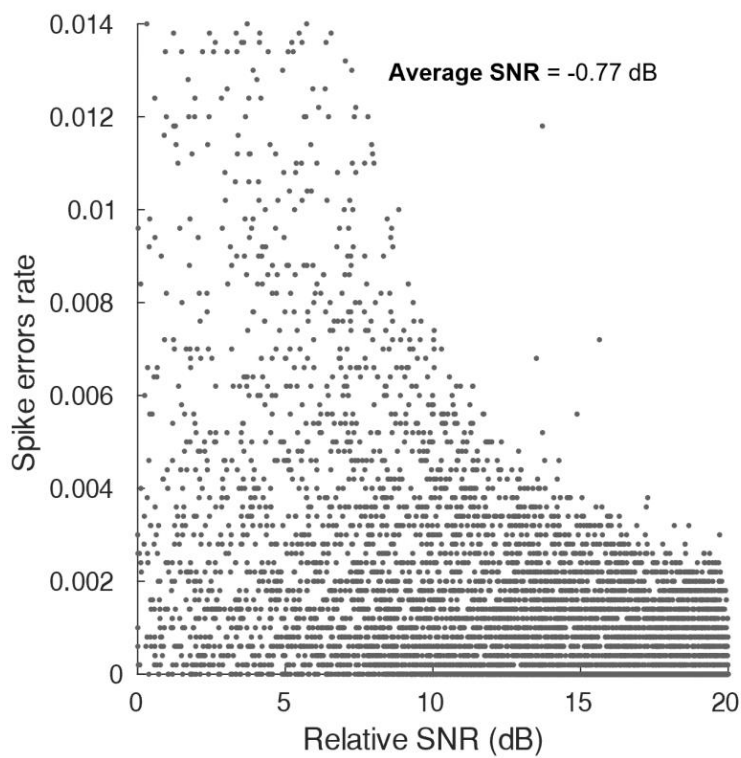

**Supplementary Figure 8.** The spike error rate (SER) in transmitting 8,200 nodes of neural spike data. The simulation addresses the near-far problem by varying the amplitude of backscattering which leads to differences in relative SNR (0 to 20 dB) among chips. The results indicate that nodes transmitting weaker signals, those located farther away from the external receiver, exhibit higher spike errors compared to those generating stronger backscattering as they are located closer to the external receiver.

| Table S2 | Correlation analysis between the original x-velocity of the cursor and its reconstructed value using a neural decoder | | | | |
|---|---|---|---|---|---|
| Content | Correlation (5-folds) | | | | |
| M1 (collected in a wired system) | 0.93911 | | | | |
| M1 (transmitted through the ASBIT protocol, SNR= -12.77) | 0.658621 | 0.720138 | 0.752318 | 0.732099 | 0.717825 |
| M1 (transmitted through the ASBIT protocol, SNR= -8.77) | 0.794375 | 0.854243 | 0.874156 | 0.804641 | 0.843862 |
| M1 (transmitted through the ASBIT protocol, SNR= -4.77) | 0.878084 | 0.900917 | 0.905043 | 0.875753 | 0.89279 |
| M1 (transmitted through the ASBIT protocol, SNR= -0.77) | 0.913785 | 0.924004 | 0.929386 | 0.903682 | 0.921177 |
| M1 (transmitted through the ASBIT protocol, SNR= 3.23) | 0.926512 | 0.936714 | 0.942347 | 0.923157 | 0.935254 |
| M1 (transmitted through the ASBIT protocol, SNR= 19.23) | 0.922885 | 0.936211 | 0.943901 | 0.922071 | 0.936769 |
| | | | | | |
| S1 (collected in a wired system) | | | | | |
| S1 (transmitted through the ASBIT protocol, SNR= -12.77) | 0.671793 | 0.690964 | 0.671976 | 0.643869 | 0.58587 |
| S1 (transmitted through the ASBIT protocol, SNR= -8.77) | 0.808395 | 0.809847 | 0.808138 | 0.799064 | 0.788678 |
| S1 (transmitted through the ASBIT protocol, SNR= -4.77) | 0.874804 | 0.873204 | 0.873599 | 0.863752 | 0.859218 |
| S1 (transmitted through the ASBIT protocol, SNR= -0.77) | 0.903724 | 0.902957 | 0.910874 | 0.900415 | 0.897612 |
| S1 (transmitted through the ASBIT protocol, SNR= 3.23) | 0.922929 | 0.920866 | 0.924606 | 0.923442 | 0.913239 |
| S1 (transmitted through the ASBIT protocol, SNR= 19.23) | 0.928275 | 0.932218 | 0.933293 | 0.932113 | 0.922434 |

**Supplementary Note 6. Power and loss considerations for building a wireless power transfer system for multisensory neural interface.**

Although our work mainly demonstrates the communication idea in the ASBIT method, it is important also to anticipate possible limitations in wireless power transfer efficiency in building wireless neural or other body-implanted sensor networks. As explained in Supplementary Note 2, our present communication microchip requires a minimum power of 25.11 µW (-16 dBm), and a wireless transfer efficiency of -40 dB is necessary for chip operation at 24 dBm transmitting power. Although power consumption can be improved through smaller nodes or better ASIC circuit designs, and is not close to fundamental limits, we used these parameters to assess the feasibility of a wireless neural sensor interfaces with thousands of distributed devices.

As an example of a possible configuration, we simulated a 4-coil wireless link, including our previously reported approach [2] now with a 300 µm × 300 µm square microcoil in the perimetry of each sensor. The first relay coil was assumed to be placed 5 mm away from the Tx external transmitting coil (2 mm of skin and 3 mm of fat layers), while the second relay coil and the on-chip microcoil were coplanar and located 7 mm apart from the first relay coil (7 mm of the skull), as illustrated in Supplementary Fig. 9a. For epicortical applications, we assumed that electrodes attached to individual chips penetrate the cortex to access neural spikes. We then used Ansys, HFSS (high-frequency structure simulator) to solve for the quality factor of the coils and the coupling factor between them, and calculated the wireless transfer efficiency in the 4-coil system as described in [21]. The geometry of the Tx and microcoil on-chip, as well as the simulated efficiency in 2,025 locations spaced 300 µm apart within one quadrant of the transmitting coil, is shown in Supplementary Figure 9. (As a practical matter, the relay coils, deposited on flexible polymer substrates a fraction of a millimeter in thickness, are relatively simple to insert into tissue and have been used by us in rodent implants [2]).

Based on our simulations, we found that the wireless transfer efficiency between our the Tx coil and on-chip Rx microcoils ranged from -37.3 dB to -29.3 dB, resulting in power received on the chips ranging from 46.7 µW to 295.1 µW. All locations within one quadrant of the transmitting coil had an efficiency above -40 dB and each quadrant of the relay coil covers an area that can accommodate up to 900 microchips. With four symmetrical quadrants, up to 3,600 microchips can be placed within the Tx coil in this design, allowing for the simultaneous collection of spike signals from thousands of chips. However, the impedance mismatch between the coil and circuit and loading effects from other microchips can decrease the wireless transfer efficiency. Therefore, the current circuit design might require more than allotted 24 dBm of total Tx power, or a sophisticated resonance tuning mechanism to minimize the loading effects. Another path is to pursue the development of wireless sensors with lower power requirements than the -16 dBm number above, leveraging semiconductor technology at an advanced process beyond the 65 nm CMOS node. To achieve an 8,000 neural sensor network, as simulated in Fig.4 of the main text, additional choices can involve the use of multiple Tx external antennas [2] to distribute incident RF power across a wider area while remaining below SAR limits.

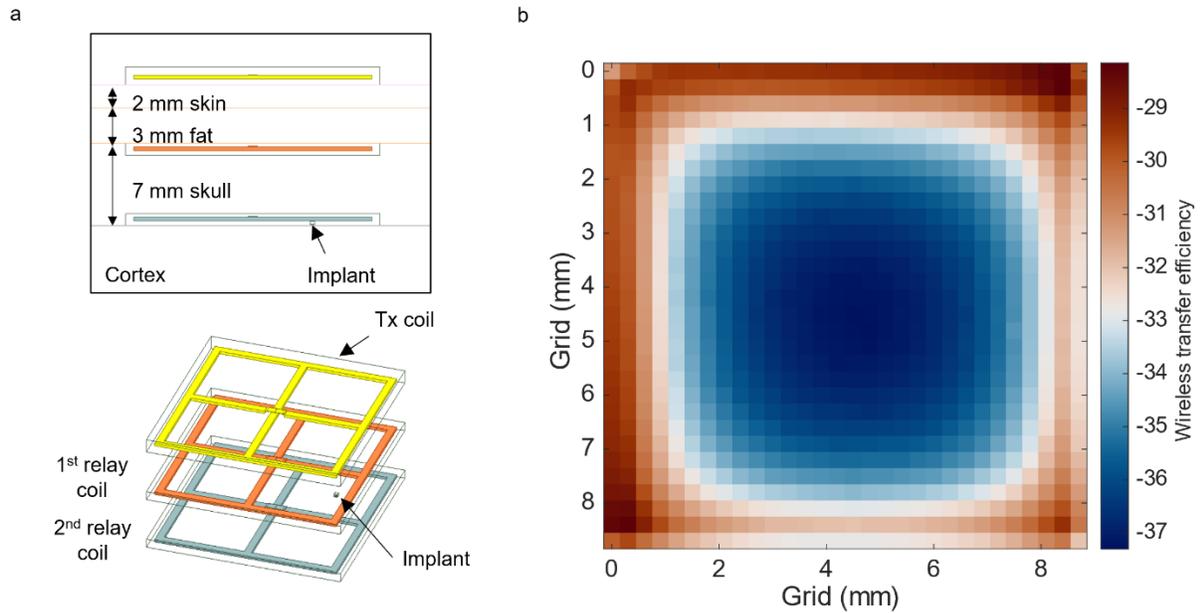

**Supplementary Figure 9.** Possible wireless energy transfer configuration for a large-scale neural interface. a) Schematic cross-section of a 4-coil system geometry for a wireless neural interface composed from several thousand microsensors. The system includes one external Tx coil (20 mm × 20 mm), two implanted relay coils (20 mm × 20 mm), and an on-chip microcoil on each implanted sensor chip (300 µm × 300 µm in area). The Tx coil is positioned over the skin, while the first and second sub-mm thick relay coils are inserted above and below the skull, respectively. Further details on the suggested coil design can be found in [2]. b) Heatmap of the simulated wireless transfer efficiency across the population of sensor chips for one quadrant of the 'window coils'. The results for the other three quadrants are symmetrical, and therefore only one quadrant is shown in the image. This plot presents that the wireless power transfer efficiency ranges from -37.3 dB to -29.3 dB in this 4-coil system. Each quadrant of the relay coil can accommodate up to 900 closely spaced microchips.

**References:**

[1] Kumar, V. Anil, Abhijit Mitra, and SR Mahadeva Prasanna. "On the effectivity of different pseudo-noise and orthogonal sequences for speech encryption from correlation properties." International Journal of Electronics and Communication Engineering 2.12 (2008): 2844-2851

[2] Lee, J. *et al*. Neural recording and stimulation using wireless networks of microimplants. *Nature Electronics* **4**, 604–614 (2021).

[3] Determining the peak spatial-average specific absorption rate (SAR) in the human body from wireless communications devices, 30 MHz to 6 GHz - part 3: Specific requirements for using the finite difference time domain (FDTD) method for SAR calculations of mobile phones. IEC/IEEE 62704-3:2017 1–76 (2017).

[4] Cao, Y., Zhang, L., Chang, C.-H. & Chen, S. A low-power hybrid ro PUF with improved thermal stability for lightweight applications. *IEEE Transactions on computer-aided design of integrated circuits and systems* **34**, 1143–1147 (2015).

[5] Proakis, J. G., Salehi, M., Zhou, N. & Li, X. *Communication systems engineering*, vol. 2 (Prentice Hall New Jersey, 1994).

[6] Gilhousen, K. S. *et al*. On the capacity of a cellular CDMA system. *IEEE transactions on vehicular technology* **40**, 303–312 (1991).

[7] Turkmani, A. M. & Goni, U. Performance evaluation of maximal-length, Gold and Kasami codes as spreading sequences in CDMA systems. In *Proceedings of 2nd IEEE International Conference on Universal Personal Communications*, vol. 2, 970–974 (IEEE, 1993).

[8] Klair, D. K., Chin, K.-W. & Raad, R. A survey and tutorial of RFID anti-collision protocols. *IEEE Communications surveys & tutorials* **12**, 400–421 (2010).

[9] Eom, J.-B., Yim, S.-B. & Lee, T.-J. An efficient reader anticollision algorithm in dense RFID networks with mobile RFID readers. *IEEE Transactions on industrial electronics* **56**, 2326–2336 (2009).

[10] Hush, D. R. & Wood, C. Analysis of tree algorithms for RFID arbitration. In *Proceedings. 1998 IEEE International Symposium on Information Theory (Cat. No. 98CH36252)*, 107 (IEEE, 1998).

[11] Myung, J., Lee, W. & Srivastava, J. Adaptive binary splitting for efficient RFID tag anti-collision. *IEEE communications letters* **10**, 144–146 (2006).

[12] Park, J., Chung, M. Y. & Lee, T.-J. Identification of RFID tags in framed-slotted ALOHA with robust estimation and binary selection. *IEEE Communications Letters* **11**, 452–454 (2007).

[13] Liva, G. Graph-based analysis and optimization of contention resolution diversity slotted ALOHA. *IEEE Transactions on Communications* **59**, 477– 487 (2010).

[14] Eom, J.-B. & Lee, T.-J. Accurate tag estimation for dynamic framed-slotted ALOHA in RFID systems. *IEEE Communications Letters* **14**, 60–62 (2009).

[15]     Yeon, P., Bakir, M. S. & Ghovanloo, M. Towards a 1.1 mm2 free-floating wireless implantable neural recording SoC. In 2018 IEEE Custom Integrated Circuits Conference (CICC), 1–4 (IEEE, 2018).

[16]     Ahmadi, N. et al. Towards a distributed, chronically-implantable neural interface. In 2019 9th International IEEE/EMBS Conference on Neural Engineering (NER), 719–724 (IEEE, 2019).

[17]     Seo, D., Carmena, J. M., Rabaey, J. M., Alon, E. & Maharbiz, M. M. Neural dust: An ultrasonic, low power solution for chronic brain-machine interfaces. arXiv preprint arXiv:1307.2196 (2013).

[18] Costello, Joseph T., et al. "A low-power communication scheme for wireless, 1000 channel brain–machine interfaces." Journal of Neural Engineering 19.3 (2022): 036037.

[19] Glaser, J. I. *et al*. Machine learning for neural decoding. *Eneuro* **7** (2020).

[20] Benjamin, A. S. *et al*. Modern machine learning as a benchmark for fitting neural responses. *Frontiers in computational neuroscience* **56** (2018).

[21] RamRakhyani, Anil Kumar, Shahriar Mirabbasi, and Mu Chiao. "Design and optimization of resonance-based efficient wireless power delivery systems for biomedical implants." IEEE transactions on biomedical circuits and systems 5.1 (2010): 48-63.